\newcommand{\aeff}{{\rm A}_{\lambda}^{\rm e}}
\newcommand{\aej}{{\rm A}_{\rm J}^{\rm e}}
\newcommand{\aeh}{{\rm A}_{\rm H}^{\rm e}}
\newcommand{\aeks}{{\rm A}_{\rm K_{\rm s}}^{\rm e}}
\newcommand{\ks}{{\rm K_{\rm s}}}
\documentclass[apj]{emulateapj}
\usepackage[FIGTOPCAP]{subfigure}
\usepackage{upgreek}
\usepackage{amsmath}
\usepackage{multirow}

\begin{document}
\title{Near-Infrared Detection of a Super-Thin Disk in NGC~891}

\author{Andrew Schechtman-Rook\altaffilmark{1,2}, Matthew
  A. Bershady\altaffilmark{1}}
\submitted{Submitted to ApJ 2013 March 13; accepted 2013 June 21}
\altaffiltext{1}{University of Wisconsin, Department of Astronomy, 475
N. Charter St., Madison, WI 53706; andrew@astro.wisc.edu}
\altaffiltext{2}{Visiting Astronomer, Kitt Peak National Observatory, National
  Optical Astronomy Observatories, which is operated by the Association of
  Universities for Research in Astronomy, inc. (AURA) under cooperative
  agreement with the National Science Foundation}


\begin{abstract}
We probe the disk structure of the nearby, massive, edge-on spiral
galaxy NGC~891 with sub-arcsecond resolution $JH\ks$-band images covering
$\sim\pm$10 kpc in radius and $\pm$5 kpc in height. We measure
intrinsic surface brightness profiles using realistic attenuation
corrections constrained from near- and mid-infrared ({\it Spitzer})
color maps and three-dimensional Monte-Carlo radiative-transfer
models. In addition to the well-known thin and thick disks, a
super-thin disk with 60-80 pc scale-height---comparable to the
star-forming disk of the Milky Way---is visibly evident and required
to fit the attenuation-corrected light distribution. Asymmetries in
the super-thin disk light profile are indicative of young, hot stars
producing regions of excess luminosity and bluer
(attenuation-corrected) near-infrared color. To fit the inner regions
of NGC~891, these disks must be truncated within $\sim$3 kpc, with
almost all their luminosity redistributed in a bar-like structure 50\%
thicker than the thin disk.  There appears to be no classical bulge
but rather a nuclear continuation of the super-thin disk. The
super-thin, thin, thick, and bar components contribute roughly 30\%,
42\%, 13\%, and 15\% (respectively) to the total $\ks$-band
luminosity.  Disk axial ratios (length/height) decrease from 30 to 3
from super-thin to thick components. Both exponential and sech$^{2}$
vertical SB profiles fit the data equally well. We find that the
super-thin disk is significantly brighter in the $\ks$-band than
typically assumed in integrated SED models of NGC~891: it appears that
in these models the excess flux, likely produced by young stars in the
super-thin disk, has been mistakenly attributed to the thin disk. 
\end{abstract}

\section{Introduction}
A knowledge of the three-dimensional mass distribution of spiral galaxy disks is
critical to understanding galaxy formation and evolution. Unfortunately,
kinematic measurements of rotation curves only provide the total (stellar disk
+
dark matter)
mass distribution. Generally, this issue has been avoided by making the
assumption that stellar disks contribute all of the mass governing the
rotation curve at small radii (the ``maximal disk hypothesis'',
\citealt{vanAlbada85}). However, the recent finding of significant disk
submaximality \citep{Westfall11,Bershady11} indicates that the maximal disk
hypothesis is, at best, not universally applicable and, at worst, invalid. Additionally,
the choice of low-inclination galaxies is necessitated by the methodology of
these new mass-decomposition methods, making them insensitive to morphological
determinations of individual disk components (if multiple disks exist).

Disk mass can also be probed independently of the total gravitational
potential by measuring
the amount of starlight present in a galaxy and converting it into mass via a
mass-to-light ratio $\Upsilon_{\ast}$. Traditionally $\Upsilon_\ast$ is found
by fitting stellar population synthesis (SPS) models to the integrated galaxy
light \citep{Walcher11}. However, the values for $\Upsilon_\ast$ computed this way are dependent
on many significant assumptions, leading to large uncertainties in
$\Upsilon_\ast$ (see \citealt{Bershady10a} for a discussion). Additionally, any
estimates of luminosity are dependent on attenuation from dust within the
galaxy, which is known to be both clumpy and inclination dependent (see
\citealt{Schechtman-Rook12b} and references therein). Lastly, as these SPS
models are created with the integrated light of a galaxy, they cannot
distinguish between different stellar disks, the detailed parameters of which
can be significant for estimates of the disk mass
\citep{Bershady10a,Bershady10b}.

Observing edge-on galaxies presents the best conditions for the identification
and characterization of multiple disks, as any projection effects are
minimized. However, the effects of dust attenuation are also maximized in
galaxies with this orientation. While not as significant an issue in low-mass
galaxies, where the dust is not tightly concentrated into a narrow lane
\citep{Dalcanton04} and some galaxies are within the {\it Hubble Space
  Telescope}'s range for resolved stellar population studies of the vertical
structure \citep{Seth05}, in high mass galaxies similar to the Milky Way (MW)
determining disk structure near the midplane is a considerable challenge. 

The
problem of attenuation from the dust lanes of massive, edge-on galaxies is
mitigated somewhat by observing in the infrared, where light is minimally
absorbed by dust. Ideally observations would be taken in the mid-infrared (MIR),
around 3-4 $\mu$m, where attenuation is at its minimum and just blueward of the
point where thermal emission from cool dust begins to dominate the spectral energy
distribution (SED). However, this spectral region contains several narrow
emission lines from non-thermal polycyclic aromatic hydrocarbon (PAH)
molecules, as well as some continuum flux from very hot, small grains
\citep{Meidt12}. Additionally, to avoid the high opacity of the Earth's
atmosphere at these wavelengths, observations in the MIR must be taken on
space-based telescopes, which places strict limitations on the size of their mirrors. Current-generation space-based
telescopes with MIR capabilities, such as the {\it Spitzer Space Telescope}
\citep{Werner04} and {\it Wide-field
Infrared Survey Explorer} \citep{Wright10}, have relatively poor
($>$1\arcsec) resolution, which is generally too coarse to resolve surface-brightness
profiles very near the midplane in all but the closest galaxies (90 pc,
roughly the scale height of the young disk in the MW \citep{Bahcall80}, is
$<$1\arcsec at 20 Mpc). At these
wavelengths the diffraction limit also starts to become a significant limiting
factor on resolution: {\it Spitzer}, for example, has a diffraction limit of
$\sim$1.2\arcsec at 4$\mu$m.

Ground-based near-infrared (NIR; $\sim$0.7-2.3$\mu$m) detectors have improved dramatically in recent
years and now can rival optical detectors in resolution and noise
characteristics (e.g. \citealt{Kissler-Patig08,Meixner10}). However, while PAH emission is no longer a concern,
attenuation can still be significant; \citet{Xilouris99} find a central
edge-on K-band optical depth of $\sim$3 for NGC~891, for example. Due to the
higher resolution and lack of dust emission, however, the NIR is the best choice for detailed studies of
stellar disk parameters near the midplane of massive galaxies. 

Other researchers have turned to the NIR in order to reduce the dust-lane
attenuation of edge-on spirals. \citet{Barnaby92} fit the surface-brightness
distribution of an H-band image of NGC~5907 with a single vertical disk component, while \citet{Rice96} perform a
similar fit to NGC~4565 in the K-band. Both studies follow \citet{Wainscoat89}
in assuming that even for edge-on galaxies the effects of dust attenuation in
the NIR are minimal\footnote{\citealt{Barnaby92} do make a small attempt to
include some dust effects.}, which may contribute to their difficulties in fitting the
vertical profiles. More recently, \citet{Yoachim06} fit 2D models to edge-on
galaxies from a survey with FWHM resolution of $\sim$1\arcsec in the $\ks$-band
\citep{Dalcanton00}. However, this survey is not local enough to resolve super-thin disks (1\arcsec$\sim$300 pc at
the sample's average Hubble flow distance of 67 Mpc) and is also deliberately
biased towards low surface-brightness galaxies.

\citet{Aoki91} do make an effort to correct for the effects
of dust in K-band vertical surface brightness profiles of NGC~891. However,
they model the dust as a foreground screen, as opposed to a more physically
realistic (albeit more complex) intermixture of dust and
starlight. \citet{Xilouris99} use a smooth. axisymmetric distribution of dust
and starlight to fit radiative transfer models to optical and infrared data on a small sample of nearby
edge-on galaxies, including NGC~891. While
comparable to or better than current space-based capabilities in the MIR, none of these studies have high enough
resolution in the NIR to fully explore any disk components with scale-heights
similar to the MW's young, star-forming disk at heights where they are likely
to dominate the vertical surface-brightness profile. 

In this paper we present high-resolution ($< 1\arcsec$) NIR observations of NGC~891, a massive
edge-on spiral galaxy often compared to the MW. As a relatively isolated
system without significant signs of interaction in its starlight (although it
contains significant amounts of extra-planar gas; \citealt{Oosterloo07}), NGC
891 is an excellent test case for studies of passive disk evolution in massive
galaxies. 

We combine these NIR observations with 3-dimensional Monte Carlo radiative
transfer (RT) models, based on prior studies of NGC~891
\citep{Schechtman-Rook12b,Schechtman-Rook12c}, which employ fractally clumped
dust and dust-enshrouded star-formation. We use these models to compute an
attenuation correction that is both astrophysically realistic and specifically
tailored to NGC~891. In order to obtain a wider color baseline for the
attenuation correction, we use archival {\it Spitzer} IRAC \citep{Fazio04} 3.6 and 4.5$\mu$m data in
complement to the NIR data. While these data have a larger pixel size than the
NIR data the surface brightness is generally smoothly varying, and as such
introduces only a small amount of error into the attenuation corrections while
allowing for a much tighter correction even at high attenuation.

We then apply this
correction to our data, allowing us to produce fully attenuation-corrected
surface-brightness (SB) profiles of NGC~891 and for the first time investigate
disk morphology near the midplane of a massive spiral galaxy outside of the MW.

This paper is organized as follows: In Section \ref{sec:data} we present our data
and describe our reduction software. We then detail our approach to removing
the attenuation in Section \ref{sec:dustcorr}. Light profiles and
two-dimensional fits to the data are presented in Section
\ref{sec:sbprofiles}. We discuss our results in a broader context in Section
\ref{sec:discussion}, and make concluding remarks in Section
\ref{sec:conclusion}. We assume a distance to NGC~891 of 9.5 Mpc, so 1'' = 46 pc.

\begin{figure*}
\begin{centering}
\plotone{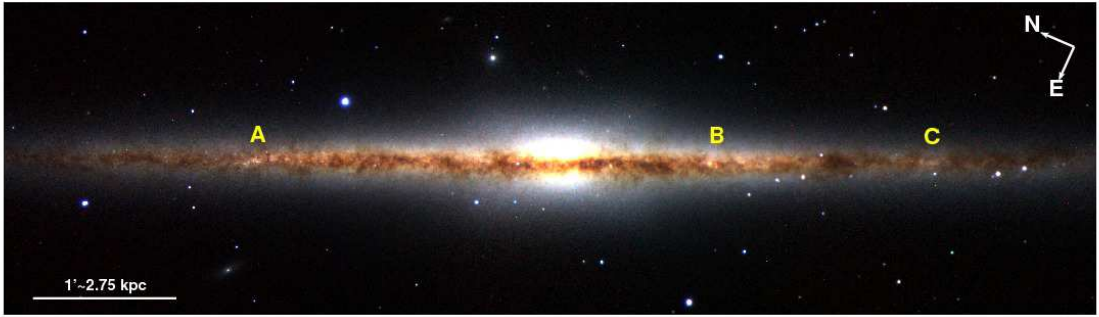}
\caption{False-color RGB image of NGC~891, using the J (blue), H (green), and $\ks$
  (red) WHIRC data analyzed here. Labels (A, B, and C) are placed directly
  above three regions which appear to have enhanced, localized star formation on
  the side of the galaxy nearest to us. At the assumed distance, 1\arcmin
  corresponds to $\sim$2.75 kpc. The image has a width of 7\farcm 7$\sim$21.2
  kpc and a height of 2\farcm 2$\sim$6.1 kpc. The image is centered at
  2:22:33.14,+42:20:57.5, and has a position angle of 66.7$^{\circ}$.}
\label{fig:jhkimage}
\end{centering}
\end{figure*}
\section{Near Infrared Imaging Data}
\label{sec:data}
\subsection{Ground-Based Observations}
All observations were obtained with the WIYN\footnote{The WIYN observatory
  is a joint facility of the University of Wisconsin-Madison, Indiana
  University, Yale University, and the National Optical Astronomy
  Observatory.} 3.5 m telescope at Kitt Peak
National Observatory during an observing run in October 2011. We used the
WIYN High-resolution InfraRed Camera (WHIRC; \citealt{Meixner10})
near-infrared imager, and took data in the broadband J, H, and $\ks$ filters. WHIRC has a
3\farcm3 field-of-view and 0\farcs1 pixel scale, which allows it to
fully exploit the excellent seeing at WIYN (median seeing $\sim$0\farcs7). 

\begin{deluxetable}{c c c c c c}
\tablewidth{0pt}
\tablecaption{Log of Observations}
\tablehead{\colhead{} & \colhead{} & \colhead{} & \multicolumn{2}{c}{Exposure
    Time (s)} & \colhead{}\\
\colhead{Filter} & \colhead{Date (UT)} & \colhead{Region} &
\colhead{Individual} &\colhead{Total} & \colhead {Seeing FWHM}}
\startdata
J...&2011 Oct 18&SW&80&640&0\farcs 6\\
 &2011 Oct 19&Center&80&640&0\farcs 9\\
 &2011 Oct 19&NE&80&640&0\farcs 9\\
H...&2011 Oct 18&SW&80&640&0\farcs 8\\
 &2011 Oct 19&Center&80&640&0\farcs 7\\
 &2011 Oct 19&NE&80&640&0\farcs 5\\
$\ks$...&2011 Oct 18&SW&40&1080&0\farcs 9\\
 &2011 Oct 19&Center&40&1080&0\farcs 5\\
 &2011 Oct 19&NE&40&1080&0\farcs 5\\
\enddata
\label{tab:obslog}
\end{deluxetable}

A log of the
observations is presented in Table \ref{tab:obslog}. Three regions of NGC~891
(the center, along the major axis NE of the center, and along the major axis
to the SW) were observed to provide coverage on both sides of the galaxy out
to $\sim$ 2.5 and 14 K-band thin disk scale-lengths and scale-heights, respectively (10
and 4.7 kpc; \citealt{Xilouris99}). Enough
overlap was left between these pointings to allow for combined mosaics to be
produced. Multiple
exposures were taken in each band at each pointing in a dither pattern (with
offsets between dithers $\sim$ 15\arcsec for J and H and 25\arcsec for $\ks$) and
averaged into a mosaic; the
seeing estimates provided in Table \ref{tab:obslog} are for the image mosaics and come from Gaussian fits to foreground stars in the image using the
IRAF\footnote{IRAF is distributed by the National Optical Astronomy
  Observatories, which is operated by the Association for Research in
  Astronomy, Inc., under cooperative agreement with the National Science
  Foundation.} task IMEXAM. In addition to the on-source data, we also
intersperse dithers of a sky field to aid in sky background subtraction; the
total exposure time of the sky dithers is of the same order as the total
on-source exposure time. All observations were taken under photometric
conditions. Mean seeing conditions were 0\farcs 8, 0\farcs 67, and 0\farcs 63
FWHM in J, H, and $\ks$, respectively. Full details of the data reduction can
be found in Appendix \ref{sec:rdx}, and a false-color image of our final mosaics is shown in Figure \ref{fig:jhkimage}.

\subsection{{\it Spitzer} 3.6 and 4.5 $\mu$m Data}
\label{sec:iracdata}
We downloaded ``post-basic'' calibrated data of NGC~891 from the {\it Spitzer}
archive\footnote{AOR 3632384.} in the 3.6 and 4.5 $\mu$m IRAC channels. The images were
inspected for defects. Very low level surface gradients were found across the
images and surface-subtracted using the iterative method
described in \citet{Schechtman-Rook12a} and \citet{Martinsson13}. The resolution of the IRAC data
is $\sim$1.8/arcsec (FWHM), twice as large as
the worst quality JH$\ks$ data. At these wavelengths, however, the
surface-brightness distribution of NGC~891 is fairly smooth and slowly
varying. Therefore we interpolate the IRAC data to match the WHIRC mosaics,
assuming that oversampling the IRAC images does not contribute significant
additional uncertainties to our computed attenuation correction and will not
significantly degrade the resolution of the mosaics.

\section{Dust Correction}
\label{sec:dustcorr}
Estimating the dust extinction on sight-lines to individual stars inside
the Milky Way is relatively easy because we can assume the dust is a
foreground-screen (e.g. \citealt{Cardelli89}). The task of estimating the
internal extinction 
becomes far more difficult for external galaxies due to the complex
interplay between stellar and dust geometries. For such systems the \citet{Calzetti94} obscuration
law is generally used. However, the \citet{Calzetti94} law is designed for
statistical studies of starbursts, not for detailed studies of
individual galaxies and, moreover, does not take into account the inclination
dependence of the obscuration law as seen in RT models
\citep{Matthews01, Pierini04, Schechtman-Rook12b} of disk galaxies. In the past, studies
presenting surface photometry of edge-on spiral galaxies have avoided the
midplane in their analysis
(e.g. \citealt{vanderKruit81b,Barnaby92,Barteldrees94,deGrijs96}) or used simplistic dust
corrections (e.g. \citealt{Wainscoat89,Aoki91,deGrijs97,Yoachim06}). Here, we are able
to focus on the midplane of NGC~891 using high-resolution NIR photometry and
advanced RT models.

\subsection{RT Models}
\label{sec:rtmodel}
Without knowing the 3D distribution of both dust and stars it is impossible to
perfectly attenuation-correct images of NGC~891. However, we can compute a
statistically accurate correction by modeling the dependence of color on the
attenuation at multiple points in a simulated galaxy, which produces the
average attenuation as well as the statistical uncertainty due to the varying
distributions of stars and dust in different sight-lines. This correction is
then broadly applicable for real galaxies with similar physical parameters
(e.g. dust clumpiness) and orientations. 

We base our models on the NGC~891 models of \citet{Bianchi08}, with a bulge, two stellar
disks (an old, thin disk and young super-thin disk), and a dust disk. We use
the same physical scales and un-attenuated stellar SEDs for all components as in \citet{Bianchi08}, however instead of individual molecular cloud-sized clumps
we choose to first place half of the total dust mass in fractal clumps. We
choose not to model spiral structure: the rearrangement of dust and
starlight might change the distribution of attenuation across the galaxy
but should not effect the color resulting from a given amount of attenuation. Furthermore, the addition of spiral structure has a negligible
effect on the integrated galaxy SED \citep{Popescu11}; therefore ignoring
spirality will not significantly alter the parameters of a given RT model. Both
the percentage of dust in the fractal component as well as the details of the
construction of the fractal grid follow from the median values for these
parameters found in \citet{Schechtman-Rook12b}'s modeling of NGC~891 - briefly, roughly 50\% of the
total dust mass is in clumps. In our models here the clumpy component is
divided up into 130 fractal clumps. We choose to use the
stellar emission parameters from \citet{Bianchi08} instead of this more recent work because
\citet{Bianchi08} fit the entire SED of NGC~891 whereas
\citet{Schechtman-Rook12b} were most concerned about the clumpy nature of the
dust in absorption and therefore only explored models without dust emission. The young super-thin disk
component is placed proportionately to the dust density in each grid cell. In
this way the super-thin disk is not only clumpy, but the clumpy stellar
emission is co-spatial with the dust clumps. In order to fit the integrated
SED both in the UV and the MIR, we find that we must also set a minimum dust
density threshold below which there is zero emission from young stars. This
threshold mimics the effect of having a gas surface density threshold for star formation \citep{Martin01}.

The unattenuated stellar SEDs (again, following \citealt{Bianchi08})
are set to that of an Sb stellar SED for a galaxy with no dust from
\citet{Fioc97}. The light from stars with M $<$ 3 M$_{\odot}$ is used as the
input for the bulge and thin disk, while light from stars with M $>$ 3
M$_{\odot}$ is assigned to the super-thin disk. The dust is included based on
the prescription of \citet{Draine07} with the mass fraction of PAHs
$q_{\mathrm PAH}$=4.6\%, with a size distribution to approximate
R$_{V}$=3.1 from \citet{Weingartner01}. For distribution in the model galaxy,
the dust is split into three components: large grains, with grain size $a$ $>$
200$\AA$, ultra-small (PAH dominated) grains, with $a$ $<$ 20$\AA$, and small
grains, with sizes between those of the large and ultra-small grains. These three components make up 80.63\%, 5.86\%,
and 13.51\% of the total dust mass in a given voxel (volume pixel), respectively. This combination of physical parameters we designate as Model A,
with the full set of parameters used to create the model given in Table
\ref{tab:modelparams}.

\begin{deluxetable*}{c c c c}
\tabletypesize{\footnotesize}
\tablewidth{0pt}
\tablecaption{RT Model Parameters}
\tablehead{\colhead{Parameter} &\colhead{Model A} & \colhead{Model B} &
  \colhead{Units}}
\startdata
Bulge bolometric luminosity & 1.8$\times 10^{10}$& 1.8$\times 10^{10}$&L$_{\odot}$\\
Thin disk bolometric luminosity & 4.2$\times 10^{10}$& 3.9$\times 10^{10}$&L$_{\odot}$\\
Super-thin disk bolometric luminosity & 2.8$\times 10^{10}$& 2.8$\times 10^{10}$&L$_{\odot}$\\
Bulge effective radius & 1.0& 1.0&kpc\\
Bulge axial ratio b/a & 0.6& 0.6&---\\
Thin disk scale-length & 4.0& 4.0&kpc\\
Super-thin disk scale-length & 8.0& 8.0&kpc\\
Thin disk scale-height & 0.4& 0.4&kpc\\
Super-thin disk scale-height & 0.2& 0.2&kpc\\
Dust disk central density & 3.6$\times 10^{-26}$& 3.6$\times 10^{-26}$&g cm$^{-3}$\\
Dust disk scale-length & 8.0& 8.0&kpc\\
Dust disk scale-height & 0.2& 0.2&kpc\\
Dust clumping fraction & 0.5& 0.5&---\\
Number of largest-scale clumps  &130 &130&---\\
Dust density threshold for super-thin disk & 3.4$\times 10^{-27}$& 3.4$\times
10^{-27}$& g cm$^{-3}$\\
\enddata
\label{tab:modelparams}
\end{deluxetable*}

As seen in Table \ref{tab:modelparams}, the super-thin disk of the model has a scale-length
double the size of the thin disk's scale-length but a scale-height half as
large as for the thin disk. The thin disk has an axial ratio
$\frac{h_R}{h_z}=10$, common for spiral galaxies \citep{Bershady10b}, while
the super-thin disk has a much larger $\frac{h_R}{h_z}=40$. However, due to
the minimum dust density threshold and the addition of clumpiness, the axial ratio of the super-thin disk has
less physical meaning than for the thin disk: at large distances from the
center of the galaxy and/or the midplane the super-thin disk will be
essentially truncated. The physical parameters of the thin disk and dust
component are based on the
results of \citet{Xilouris99}. \citet{Popescu11} take
a similar approach, basing dust and thin disk parameters on
\citet{Xilouris99}, but they choose a super-thin disk scale-height based on that
found in the MW. We also note that \citet{Popescu11} have an extra component
of diffuse dust with the same scale parameters as the super-thin disk.

\begin{figure}
\plotone{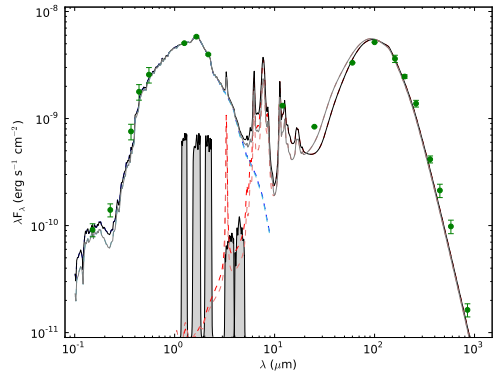}
\caption{Comparison of the integrated SEDs of the two RT
  models used in this work. Light emitted from stars (directly and scattered)
  is shown in blue (cyan), while light emitted from dust (again, both directly and
  scattered) is shown in red (pink) for Model A (B). Total emission is shown
  in black for Model A and gray for Model B. The model truncates the stellar
  emission at 10 $\mu$m; at longer wavelengths stellar emission contributes negligibly to the total flux. Green points show
observational data for NGC~891 \citep{Bianchi08}. Gray shaded regions
schematically illustrate the filter response curves of (from left to right): WHIRC J, H, and $\ks$;
IRAC 3.6$\mu$m and 4.5$\mu$m.}
\label{fig:sedmodels}
\end{figure}

For the model used here we mapped a 100x100x100 cell Cartesian grid to a
40x40x5 kpc volume, where the 5 kpc dimension represents vertical height above
and below the disk\footnote{Each cell therefore has physical dimensions of
  400x400x50 pc. For comparison, assuming a distance to NGC~891 of 9.5 Mpc, the IRAC resolution is $\sim$80 pc while the WHIRC
  resolution is $\sim$40 pc.}. We chose to employ a rectangular grid rather than a
cubical one to properly sample
the vertical extent of the dust grid, which (as shown in Table
\ref{tab:modelparams}) has a scale-height in our model of
only 200 pc. In our previous work \citep{Schechtman-Rook12b}, we used a RT
code that only tracked scattering and absorption of photons emitted at a
specific wavelength. The RT is performed by the HYPERION software package
\citep{Robitaille11}, which computes the full SED of a model. HYPERION uses an
iterative scheme to compute the dust temperature, and we use 100 million
photon packets per iteration (an average of 100 photon packets per grid cell)
for both for the temperature
calculation iterations and to produce SEDs and images. We compare the output
SED of the model to data in Figure \ref{fig:sedmodels}. While the model has
difficulty reproducing the mid-IR features associated with PAH emission and
photo-dissociation regions, \citet{Bianchi08} indicate that this is likely due
to a resolution effect in our models. Regardless of the reason for this
discrepancy we are most concerned about the SED between 1$<\lambda$ ($\mu$m)$<$5, where the model appears to
fit the data well.

\subsection{Correction Procedure}
\label{sec:correction}
To correct for the effects of dust on star-light we ran the model (described in Section \ref{sec:rtmodel}) twice: once with and once without any dust. We produced
images in J, H, $\ks$, 3.6$\mu$m, and 4.5 $\mu$m for both model runs at the inclination of NGC
891 (89.7$^{\circ}$, \citealt{Xilouris99}). The attenuation $\aeff$ at any given pixel
is then simply
\begin{equation}
\aeff=-2.5\log(\frac{F_{\lambda}}{F_{\lambda,nd}}),
\end{equation}
where $F_{\lambda}$ is the flux of the model with dust and $F_{\lambda,nd}$ is
the flux of the model without dust. Note that we use `attenuation' here broadly
to include the threefold effects of dust: absorption, scattering, {\bf and}
emission. The latter is of particular importance moving into the MIR.

As can be seen Figure \ref{fig:sedmodels} (and will be discussed again in
Section \ref{sec:multidust}), there is a very strong, narrow PAH feature which lies within the IRAC 3.6 $\mu$m bandpass. The 4.5 $\mu$m
filter, on the other hand, has no discrete dust components, although it is
somewhat affected by continuum dust emission. We find that dust
emission contributes 12\% and 9\% of the total integrated 3.6$\mu$m and 4.5$\mu$m model
flux,
respectively. Local values, along sight-lines involving dust clumps, are likely
higher: \citet{Meidt12} find ranges of 5-13\% contamination at 3.6$\mu$m in the integrated
light of {\it Spitzer} Survey of Stellar Structure in Galaxies (S$^4$G) targets but local amounts around 20\% in star-forming
regions. Therefore, based on the reduced contamination from dust emission at 4.5$\mu$m as
well as the fact that it provides a larger wavelength baseline for other
attenuation effects (absorption and scattering), we
expect that the 4.5$\mu$m data will be more useful for estimating the NIR
attenuation than the 3.6$\mu$m data. 

\begin{figure}
\plotone{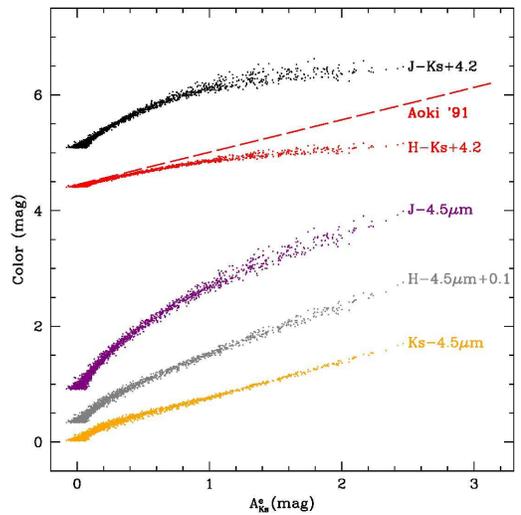}
\caption{Colors as a function of $\ks$-band attenuation for Model B. Some colors
  have been offset (noted in the labels) to improve
  clarity. Points denote
  projected model pixels at edge-on inclination. Each pixel is 0.4 x 0.05 kpc
  (radius x height), integrated over the 40 kpc line of sight in depth. The dashed red line
  shows the attenuation correction formula from \citet{Aoki91}, using H-$\ks$
  color.}
\label{fig:acorr}
\end{figure}

With these considerations in mind we inspected the relationship between
color and $\ks$-band attenuation of the model (Figure \ref{fig:acorr}). An ideal
attenuation correction metric would have both a tight and steep relationship between
attenuation and color --- i.e., a small error in
color leads to a small error in the estimated attenuation. The J- and H-$\ks$
corrections both have very small slopes at $\aeks \gtrsim 0.75$ mag, making
them poor choices for use as attenuation correctors. All NIR-IRAC colors
(NIR-4.5 $\mu$m colors are shown in Figure \ref{fig:acorr}, but the trends are very
similar for NIR-3.6 $\mu$m colors) have similar
slopes at $\aeks \gtrsim 0.75$ of $\sim$1 mag in color per mag in
attenuation, which indicates that either the attenuation in the
4.5$\mu$m band is slowly varying or dust emission is, on average, offsetting the light losses due to
attenuation at that wavelength. 

While the NIR-IRAC colors have similar responses to increasing attenuation,
the $\ks$-4.5$\mu$m (also $\ks$-3.6$\mu$m) color clearly has the least scatter in
the relationship. This is likely due to two factors: the $\ks$-band has less dust
attenuation than either J or H, and that $\ks$ is significantly redward of the
Wein peak for even the coolest stars, which makes the $\ks$-IRAC color
insensitive to the effects of different stellar populations. This feature is
especially critical because it means that the assumptions we made about the
parameters of the super-thin disk in Section \ref{sec:rtmodel} will not bias
our results. Indeed, the attenuation corrections we produce use data at all
positions in the model galaxy, and show a single trend applicable across the
entire disk. For these reasons and those given concerning MIR dust
contamination earlier in this Section, we choose to use the
$\ks$-4.5$\mu$m color to correct the attenuation. We take the 1-$\sigma$ error on the
attenuation correction in
$\aeks$ and propagate that with the errors on the data to compute the final
errors for the corrected surface-brightness images.

\subsection{Multiple Dust Distributions}
\label{sec:multidust}
\begin{figure*}
\begin{centering}
\subfigure{\includegraphics[scale=0.55]{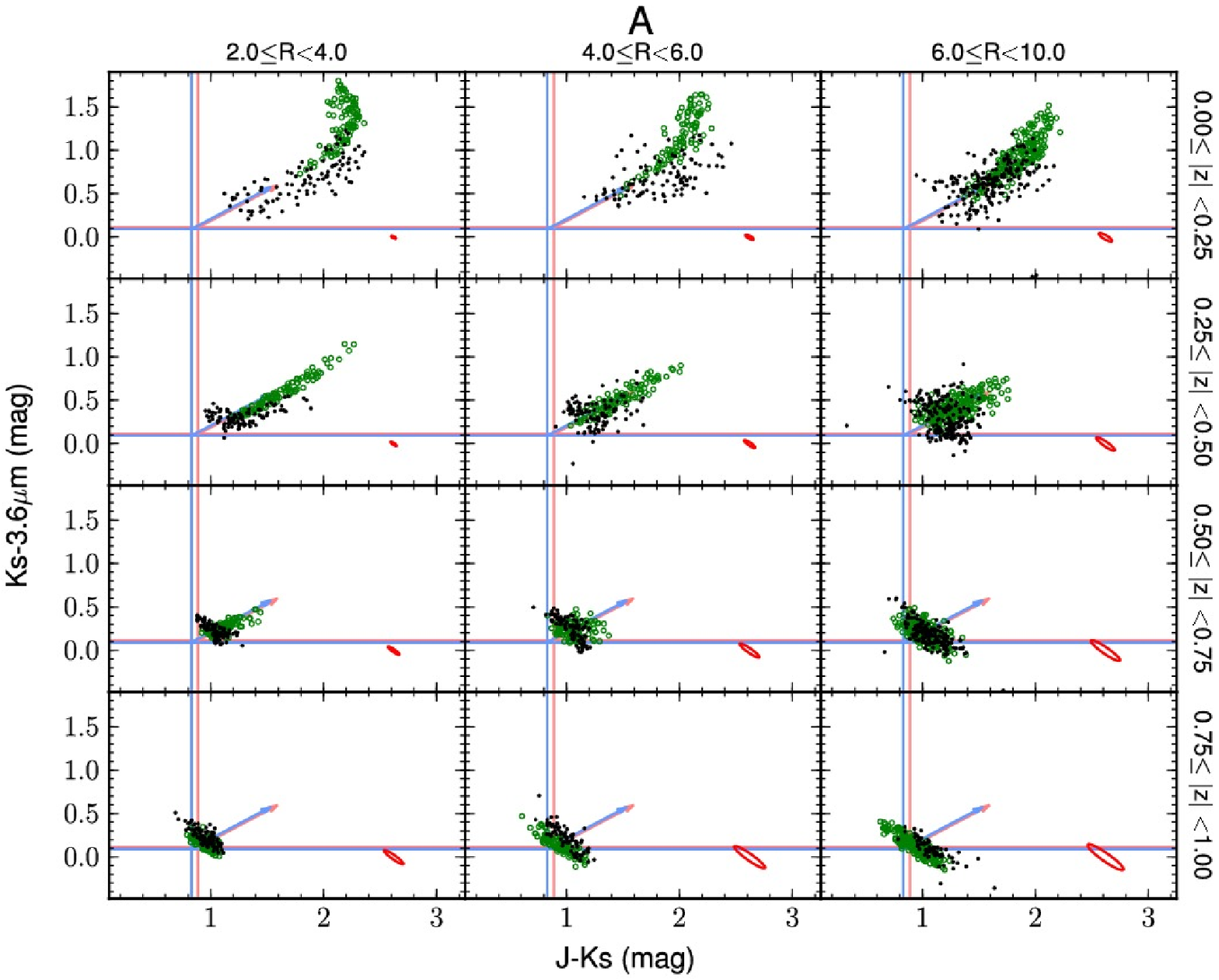}}
\subfigure{\includegraphics[scale=0.55]{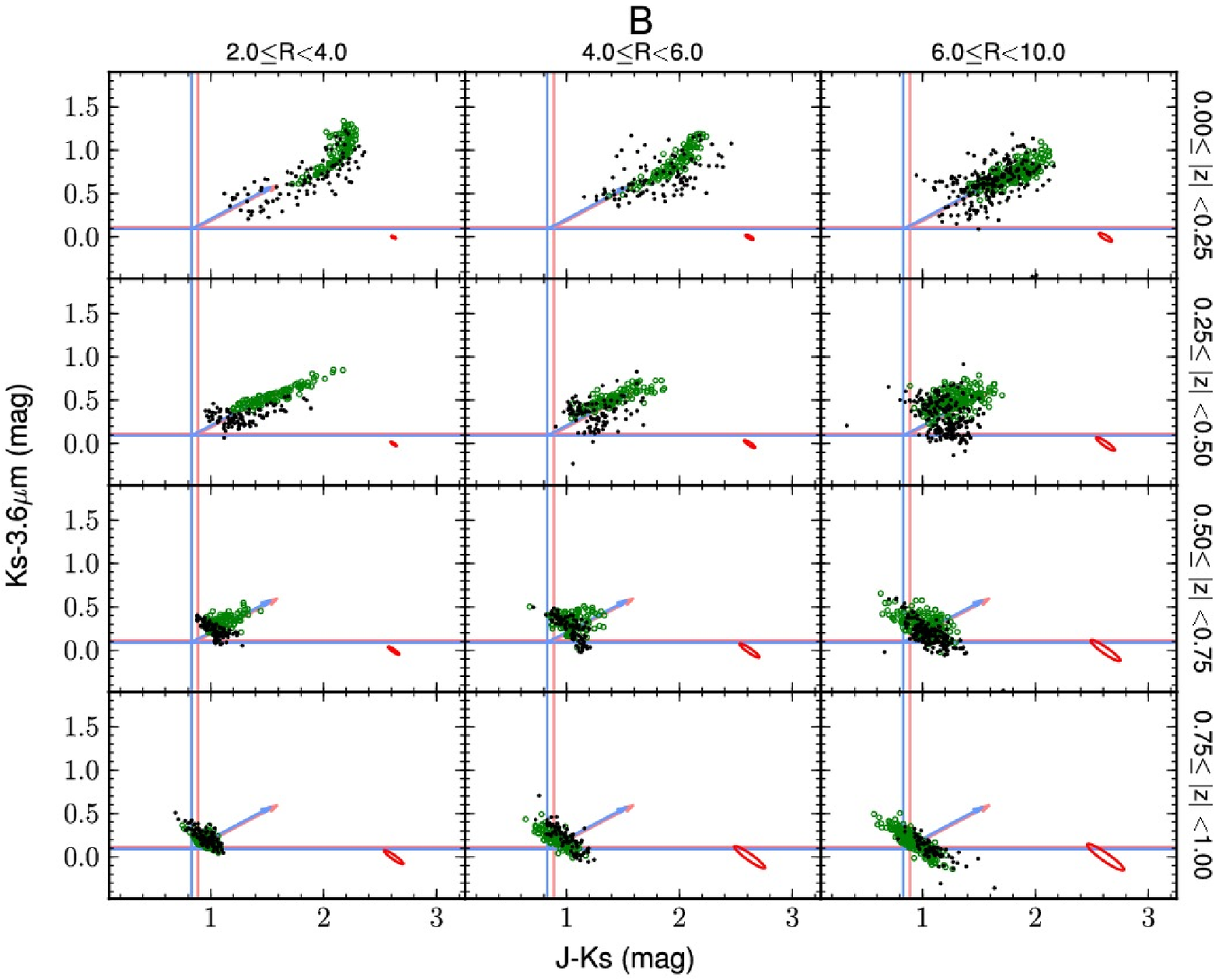}}
\caption{$\ks$-3.6$\mu$m versus J-$\ks$ color-color diagrams of Models A (top) and B
(bottom) compared to the WHIRC/IRAC data in bins of R and z (given on the top
and right axes). Model pixels are shown as open green points,
while data pixels are black filled points. The data have been resampled to
match the resolution of the model (400 x 50 pc). The red ellipses in the lower right of each
panel represent 95\% confidence
error ellipses for the data. The model (green) pixels have been perturbed by
random numbers drawn to match the error distribution in the data. The light
red (light blue) lines show the intrinsic colors of the old (young)
RT model input stellar SEDs, while the arrows approximate the reddening vector
at modest ($\aeks\sim$0.5 mag) attenuation. At $\aeks \gtrsim 1$ this vector becomes steeper
because the J-$\ks$ color index saturates before the $\ks$-3.6$\mu$m colors.}
\label{fig:cc36profiles}
\end{centering}
\end{figure*}

\begin{figure*}
\begin{centering}
\subfigure{\includegraphics[scale=0.55]{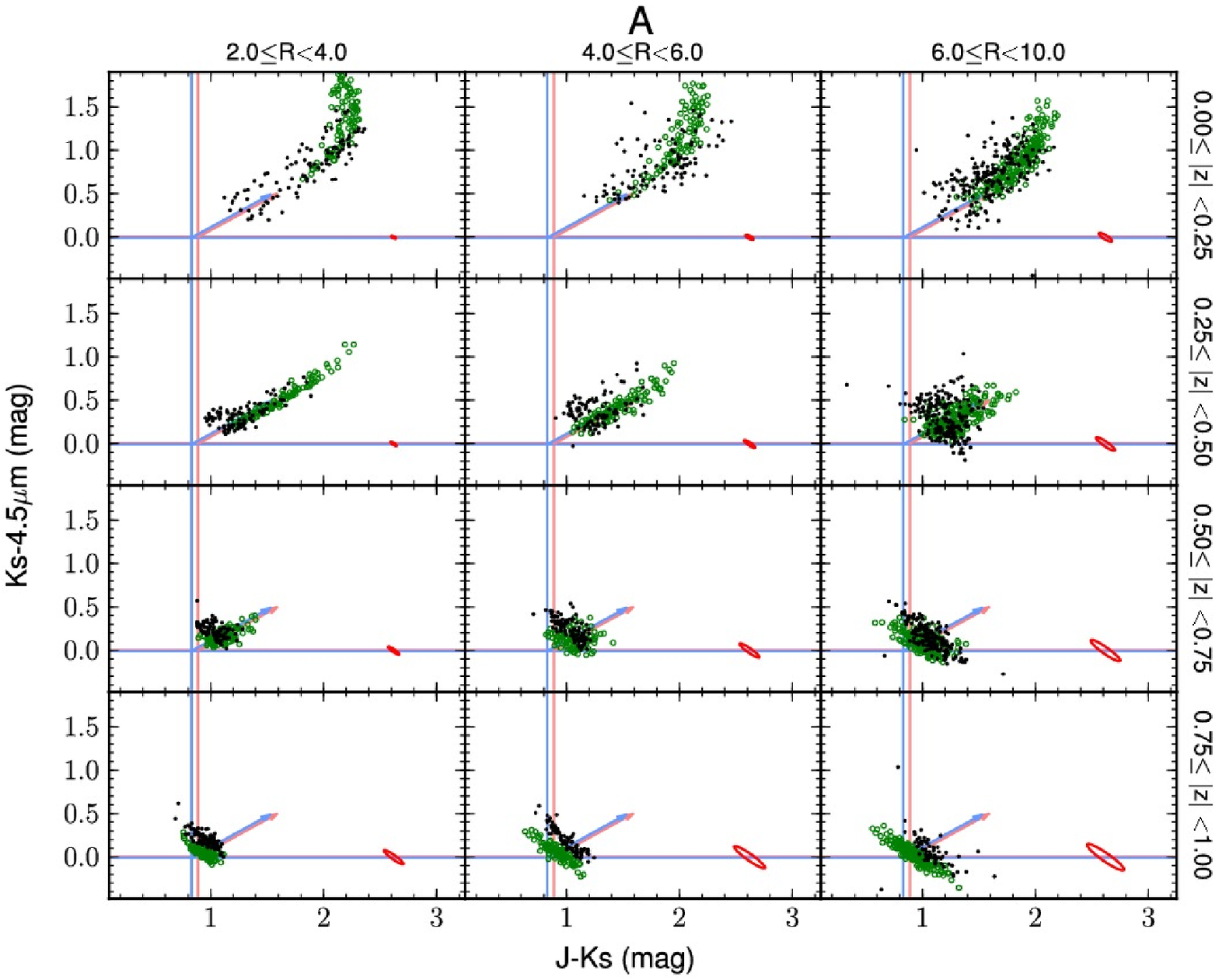}}
\subfigure{\includegraphics[scale=0.55]{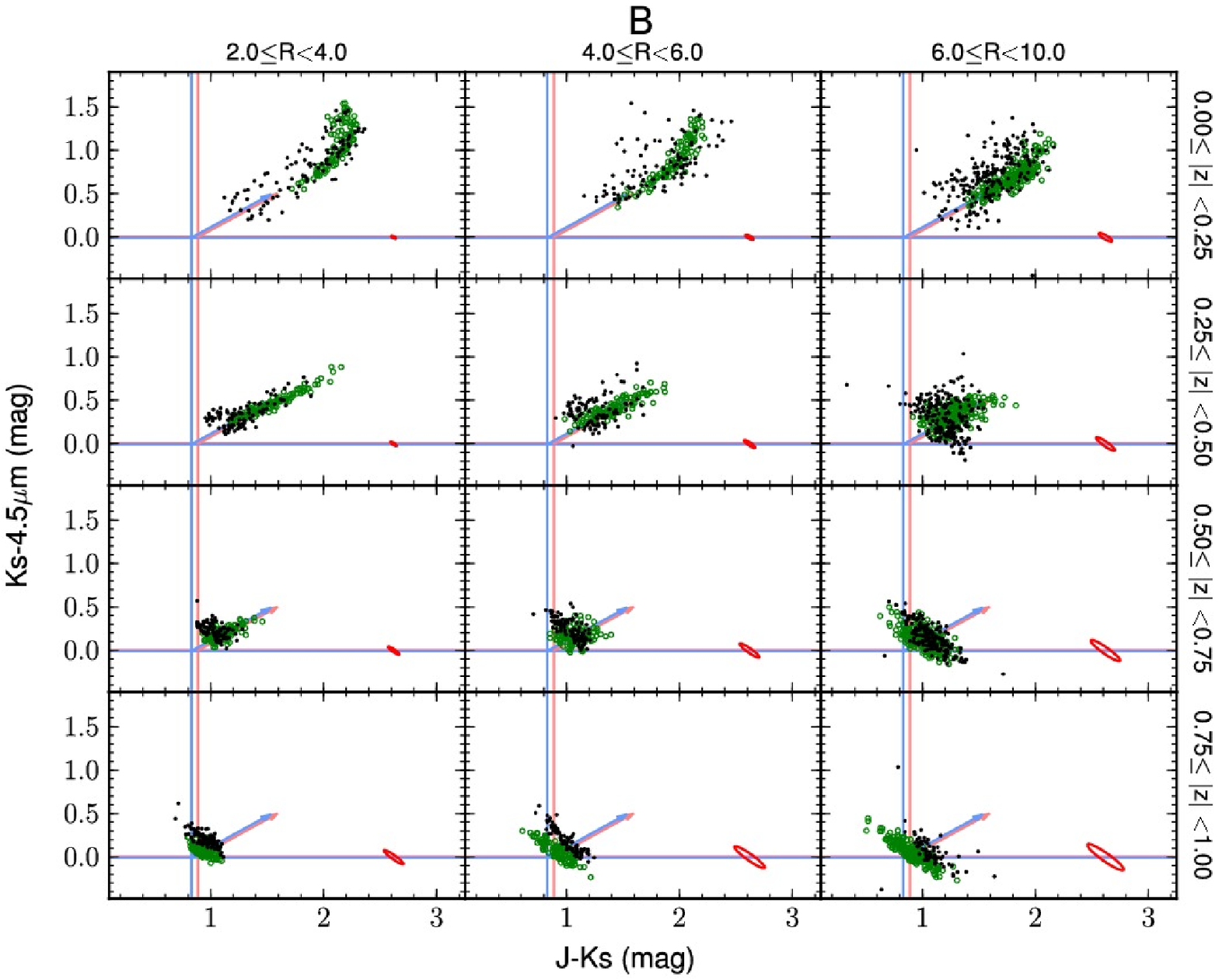}}
\caption{$\ks$-4.5$\mu$m versus J-$\ks$ color-color diagrams comparing models to
  data. Color-coding as in Figure \ref{fig:cc36profiles}.}
\label{fig:cc45profiles}
\end{centering}
\end{figure*}

To test the applicability of Model A we overlaid color-color
profiles of the model on the data and checked for discrepancies. We compared the
colors as a function of both R and z to isolate any trends with position. A
sample set of $\ks$-3.6$\mu$m versus J-$\ks$ color-color distributions is shown in Figure
\ref{fig:cc36profiles}; $\ks$-4.5$\mu$m versus J-$\ks$ distributions are shown in
Figure \ref{fig:cc45profiles}.

\begin{figure*}
\begin{centering}
\plotone{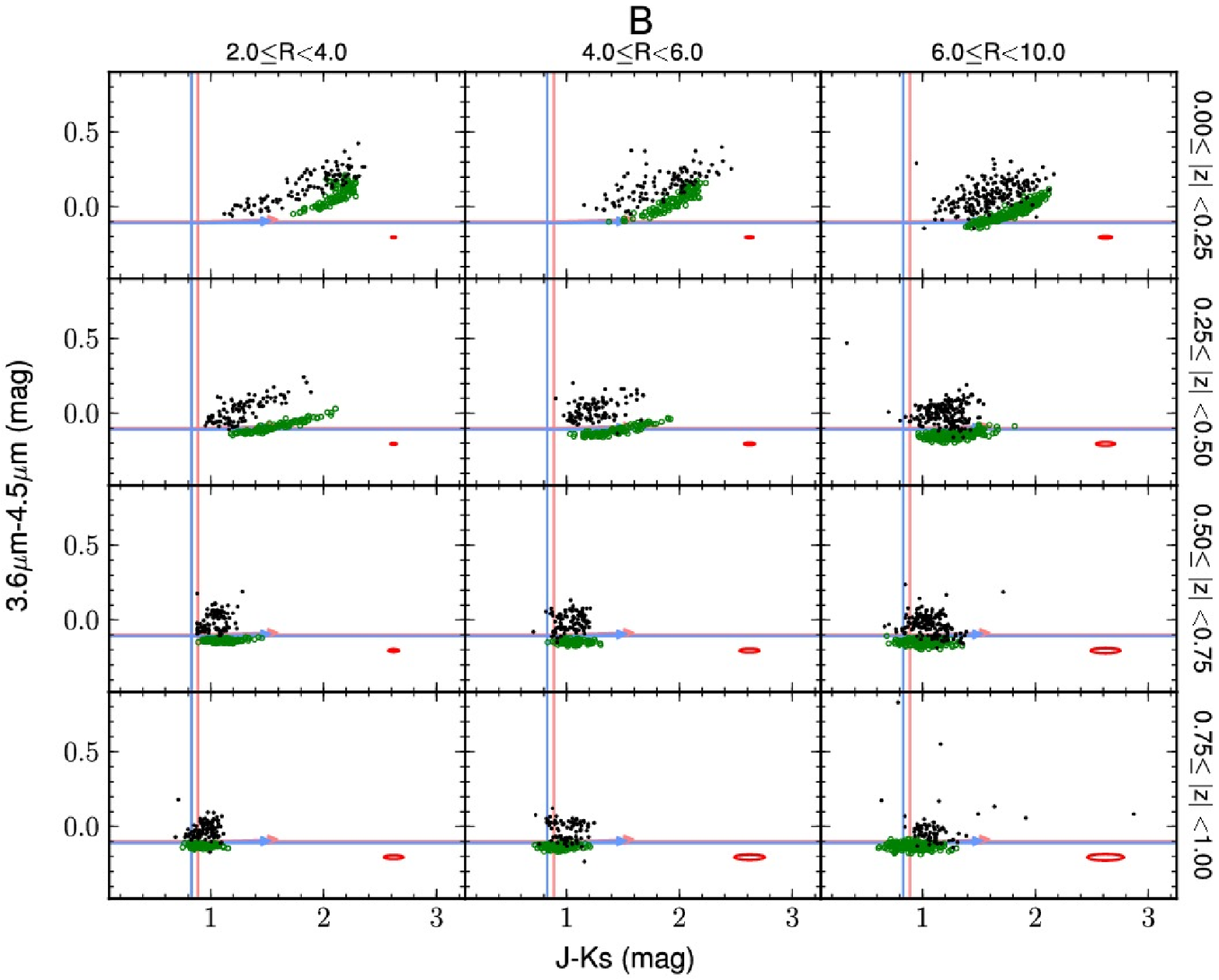}
\caption{3.6$\mu$m-4.5$\mu$m versus J-$\ks$ color-color diagrams comparing
  Model B to data. Color-coding as in Figure \ref{fig:cc36profiles}.}
\label{fig:iracccprofiles}
\end{centering}
\end{figure*}

At large distances from the midplane (high z values), both the model and the
data trend toward $\ks$-3.6$\mu$m and $\ks$-4.5$\mu$m colors of zero and J-$\ks$ colors
of $\sim$1 mag. The $\ks$-IRAC color is a result of even the
coolest stars being on the Rayleigh-Jeans tail in those spectral regions --- the
slopes of all stars are self-similar at these wavelengths, so Vega colors go
to zero. We note that the large J-$\ks$ color implies the presence of very cool
stars --- a J-$\ks$ color of 1 mag is as red as the
coldest dwarf in \citet{Bessell88}. If the NIR light at these heights is
dominated by giants, the J-$\ks$ color is roughly equivalent to a M2III star or an O-rich AGB with T$_{\mathrm{eff}}\sim$3500 K \citep{Lancon02}. 

It is also evident that at small height
Model A fails to reproduce the colors of the data; model colors are
increasingly too red in
$\ks$-3.6 and $\ks$-4.5 when z$\lesssim$0.5 kpc. Properly modeling the colors at
z$<$0.5 kpc is crucial, as the midplane is where
attenuation from dust is the strongest {\bf and} any signal from a super-thin
disk component would be most apparent. While there is better overlap when
using the $\ks$-4.5$\mu$m color (top panel of Figure \ref{fig:cc45profiles}), Model A still
produces redder $\ks$-4.5$\mu$m colors than the data.

The culprit for the discrepancy in the colors appears to be an excess of PAH
line emission in the model, as seen in Figure \ref{fig:sedmodels}: there is a
weak emission line at $\sim$3.5$\mu$m which falls within the IRAC 3.6$\mu$m
bandpass, while the wings of several lines contribute to the flux in the
4.5$\mu$m band. There are (at least) two possible reasons for this
discrepancy. The first is that there are fewer PAHs relative to other grain
sizes near the midplane of NGC~891. Reducing the amount of PAHs decreases the
total 3.6/4.5$\mu$m flux, increasing the $\ks$-3.6/4.5$\mu$m color without
affecting the J-$\ks$ color. Lowered PAH levels near the midplane could be the
result of PAH grain destruction in high density areas, similar to what is seen
in the MW (e.g. \citealt{Povich07}). On the other hand there is also
empirical evidence
that, contrary to what is frequently assumed in dust models (including the
model used here), the extinction
law flattens in the MIR \citep{Indebetouw05,Zasowski09}. Flattening the
extinction law for the IRAC bands would also have the effect of reducing the
$\ks$-3.6/4.5$\mu$m color at higher attenuations while leaving the NIR colors
unchanged. 

A full investigation of why a MW dust model has difficulty producing proper
$\ks$-3.6/4.5$\mu$m colors is beyond the scope of this work, given that we are
merely concerned with computing a proper attenuation correction rather than
conclusively determining the root cause of the attenuation. For simplicity, we
choose to adjust the PAH levels in high density regions instead of changing
the entire extinction law. Therefore we ran a second RT model with two
prescriptions for the fraction of different sized dust grains. In this new
model (called Model B) voxels with dust densities $\le$ $3.4\times 10^{-27}$ g
cm$^{-3}$ are weighted 15\% for ultra-small grains, 10\% for small grains, and 75\% for
large grains, while high-density voxels use the following breakdown of
grain sizes: 1\% ultra-small grains, 35\% small grains, and 64\% large
grains. All other quantities are held constant save the bolometric luminosity
of the thin disk, which is decreased by $\sim$7\%. These changes were made
empirically by eye to better reproduce the IR colors
of the data while still accurately fitting the integrated SED. The SED of Model B is shown alongside
that of Model A in Figure \ref{fig:sedmodels}, and color-color
$\ks$-3.6/4.5$\mu$m versus J-$\ks$ profiles are shown in the bottom panel of Figures
\ref{fig:cc36profiles} and \ref{fig:cc45profiles}. 

Model B's SED has only 9\% and 6\% of the total 3.6$\mu$m and 4.5$\mu$m flux
contributed by dust emission, generally better replicates the color
distribution of NGC~891 than Model A. There are still some minor
discrepancies between the data and the model in the MIR, however, which are
shown in Figure \ref{fig:iracccprofiles}. The model 3.6$\mu$m-4.5$\mu$m color
is systematically bluer than the the data. Near the midplane this is largely caused by the
mismatch in the 3.6$\mu$m flux as discussed earlier, but at large heights some
of the discrepancy arises from the model being slightly 
too faint at 4.5$\mu$m (visible in the bottom panels of both models in figure
\ref{fig:cc45profiles}). This mismatch is $\lesssim$0.08 mag, which implies that
the y-intercepts of the attenuation correction formulae in Appendix \ref{sec:attenuationdetails}
should be shifted by
approximately that amount. The systematic error resulting from this mismatch
will therefore be at the $\lesssim$8\% level, which should only impact
flux measurements and (as will be seen
later) does not have a significant effect on any of our conclusions. While the
3.6$\mu$m model flux does match the data at large heights, due to the
aforementioned significant mismatch in dust-obscured regions we still consider
the 4.5$\mu$m flux as more reliable for this purpose, systematic error
notwithstanding. We will
investigate ways to improve our input SEDs for future publications, but for
now we will focus on demonstrating the rest of our methodology on NGC~891. 

 Functionally, adjusting the dust distribution
decreases the reddening in the $\ks$-IRAC$\mu$m color while leaving J-$\ks$
essentially unchanged. In addition to the systematic error in intrinsic MIR
color, to have some leverage on our systematic
uncertainty caused by our imperfect knowledge of the dust distribution we
generally perform subsequent analyses with {\bf both} models,
using only Model B when we do not expect to see different responses to
analysis from the two models and the savings in computation time is significant.

\subsection{Attenuation Fits}

The exact behavior of the attenuation as a function of color will be different
in each bandpass, so we fit $\aeff$ separately for J, H, and $\ks$. Fitting the attenuation from the RT models is non-trivial,
largely because at low $\aeff$ the slope of the attenuation curve changes
significantly while at large attenuations the function becomes linear. This
problem is compounded by the dearth of model data at very high
attenuations. These factors lead to poor fits of single polynomial functions,
especially at large $\aeff$. To deal with this issue we
\begin{figure*}[h]
\begin{centering}
\plotone{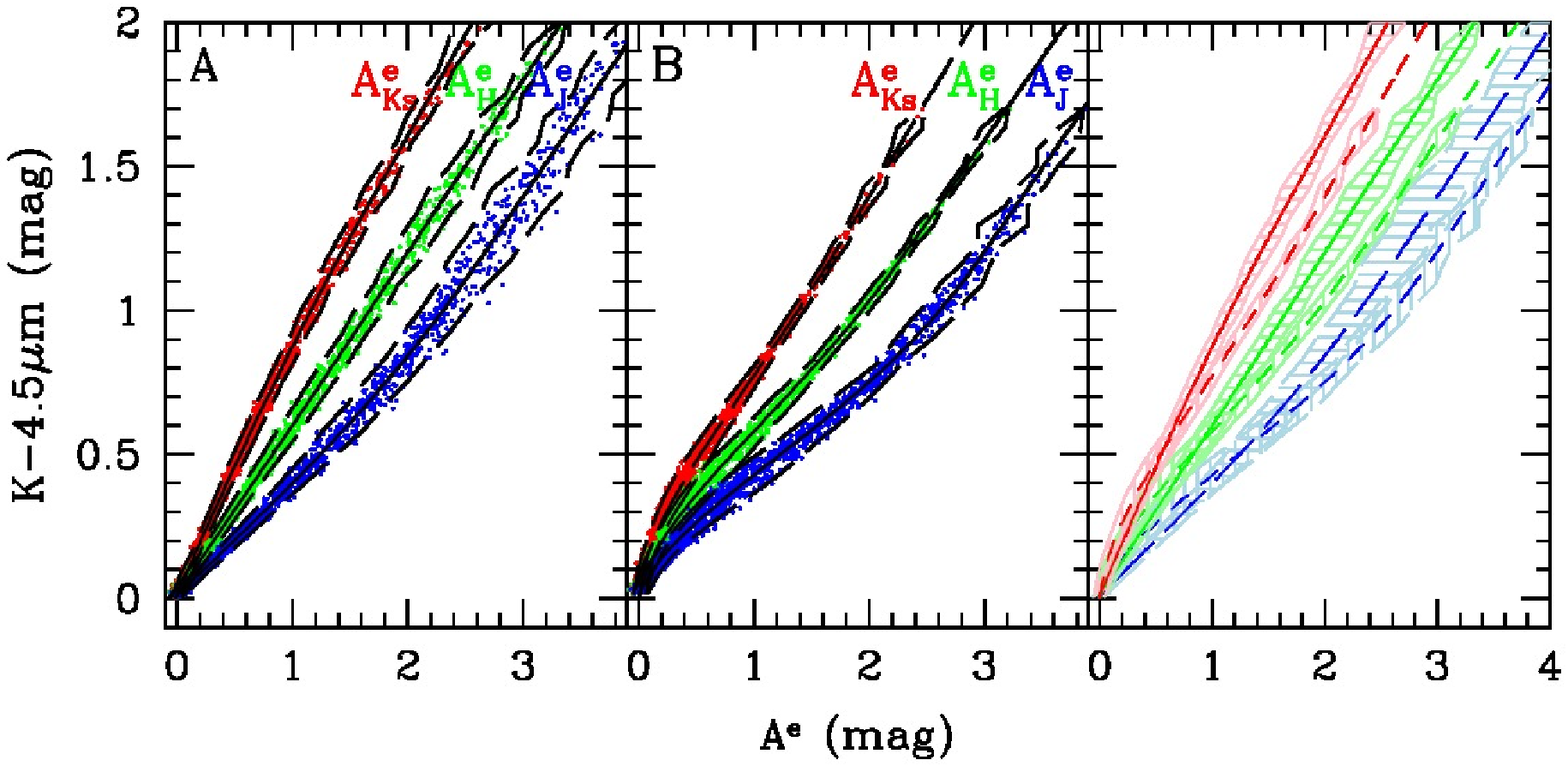}
\caption{Attenuation corrections for all three NIR bands. On the left and
  middle panels points are
  individual projected model pixels (as in Figure \ref{fig:acorr}), while the
  black solid lines show the piecewise fits to the `data'. Dashed black lines
  denote 95\% confidence limits in the distribution of $\aeff$ at a given
  model color. Left panel: Model
  A. Middle panel: Model B. The rightmost panel compares the best fitting attenuation
  correction from Model A (solid lines) and Model B (dashed lines), with
  shaded regions denoting the 2$\sigma$ errors.}
\label{fig:colorfixcompare}
\end{centering}
\end{figure*}
employ piecewise
fits of the attenuation curve. Details of our method and the formulae for the
resulting fits are shown in Appendix \ref{sec:attenuationdetails}.

The attenuation fits are plotted against the model data in Figure
\ref{fig:colorfixcompare}. In all cases it is
apparent that the model trends are well fit. In Figure
\ref{fig:colorfixcompare} we also compare the fits between Models A and B. At low
attenuations Model B
has much more complex behavior, although for all three bands the two models are within $\sim$0.1
mag of each other in $\aeff$. At large attenuations the attenuation relations have
very similar slopes but are offset from each other by about 0.3 mag in J, H,
and $\ks$. Model B has more extinction
at a given $\ks$-4.5$\mu$m color than Model A, for
$\ks$-4.5$\mu$m $\gtrsim$ 0.5 mag. The difference in behavior between the models at large values of
$\ks$-4.5$\mu$m arises because the depletion of PAH grains in Model B acts to
lower the amount of line/continuum emission from dust at $\sim$5-6$\mu$m, the tail of
which is blue enough to affect the 4.5$\mu$m IRAC band. The attenuation in the
$\ks$-band, however, is not strongly affected by changes in PAH densities and
therefore remains essentially constant. The result is that Model B has bluer
$\ks$-4.5$\mu$m colors for a given $\aeff$, seen in Figure
\ref{fig:colorfixcompare}. This also indicates that Model A overpredicts the
4.5$\mu$m surface brightness near the midplane by $\sim$0.3 mag arcsec$^{-2}$.

\subsection{Comparison with Previous Attenuation Corrections}
\begin{figure*}
\begin{centering}
  \includegraphics[scale=0.9]{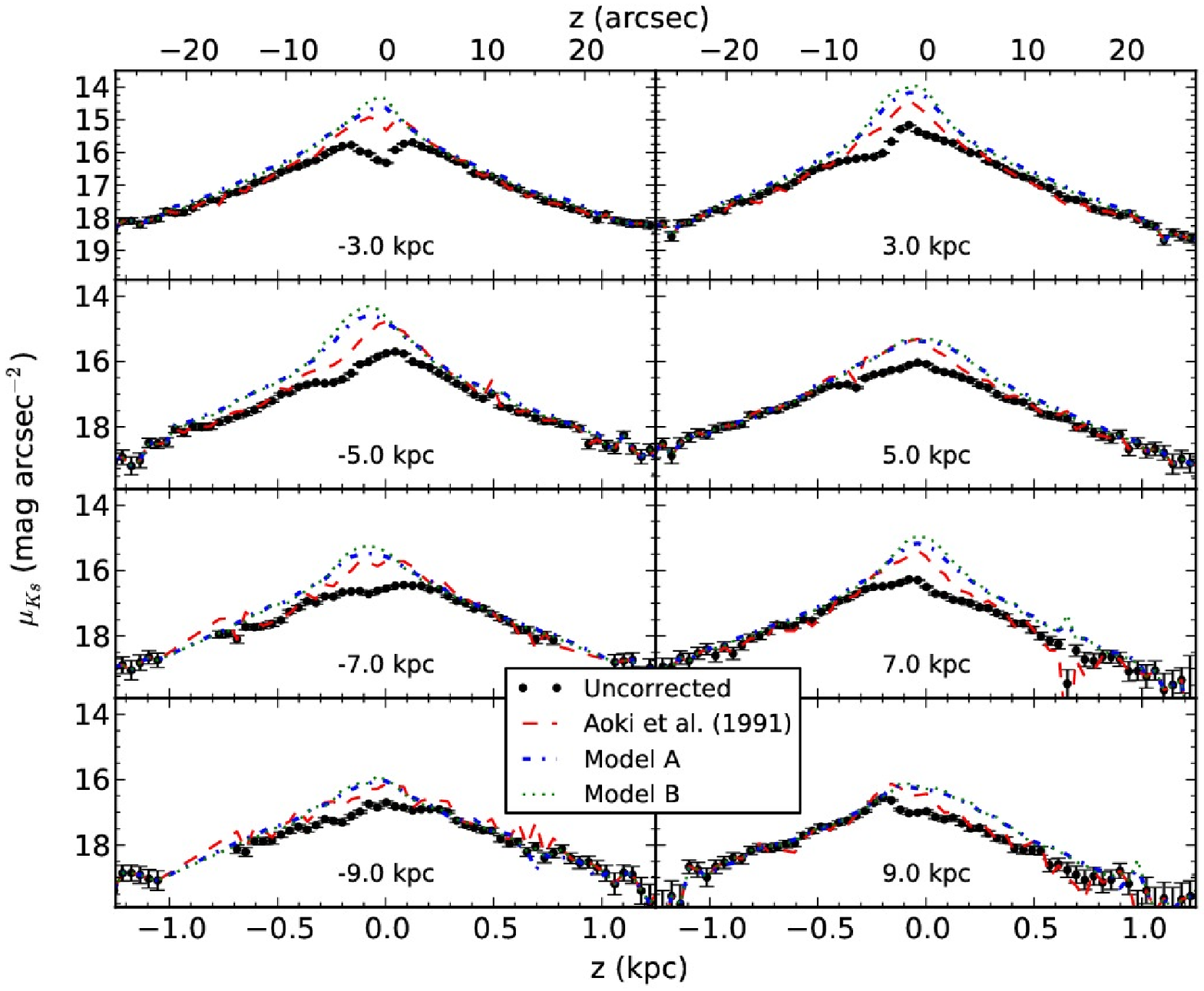}
\caption{Comparison of methods for attenuation correcting $\ks$-band data. Black
  points: uncorrected data. Red dashes: our data corrected using the foreground
  screen model of \citet{Aoki91}. Blue dot-dashed line: data corrected using
  Model A. Green
  dotted line: data corrected using Model B. The text near the bottom of each panel shows the
radius where each profile was extracted, with positive radii on the SW
side. The lack of data near $\sim$1 kpc on the -7 kpc vertical
profile and near $\sim$-1 kpc on the -7 and -9 profiles is
due to foreground stars.}
\label{fig:correctioncompare}
\end{centering}
\end{figure*}

Assuming a MW extinction law and a foreground screen of dust, \citet{Aoki91}
correct their K-band emission by the formula
\begin{equation}
K_{c}=K_{r} - \frac{(H-K)-0.25}{0.56},
\end{equation}
where $K_{r}$ is the uncorrected K-band surface brightness and $K_{c}$ is the
corrected K-band surface brightness. As the reddest filter and therefore the
data least likely to be affected by errors in the attenuation correction, we
follow \citet{Aoki91} and focus on our $\ks$-band data. We show the uncorrected $\ks$ data along
with corrected $\ks$ profiles in Figure \ref{fig:correctioncompare} on both sides
of NGC~891 at 3, 5, 7, and 9 kpc in radius. We plot the vertical profiles at these positions because
they show data at high S/N while avoiding bulge contamination, and are
representative of the profiles at all radii. While the K
and $\ks$-bands are not identical, we use the \citet{Aoki91} attenuation
correction on our data to obtain a schematic understanding of the differences
between it and our RT methodology. 

We find that generally both dust models in this work have larger peak attenuation than the
\citet{Aoki91} correction. Model A is on average brighter by $\sim$0.16 mag,
while Model B is brighter by $\sim$0.33 mag. 
 This discrepancy is likely due to the saturation of the H-$\ks$ color
at large optical depths (\citealt{Matthews01} and see Figure \ref{fig:acorr}), which we are less sensitive to since we are using
$\ks$-4.5$\mu$m color and models that accurately represent this
effect. Additionally, while the attenuation corrected vertical profiles of
both Model A and Model B peak at a relatively consistent location at all
radii, the peak of the \citet{Aoki91} dust corrected data has much more
variation (especially visible at 5 and 9 kpc). Our two RT models
are very similar to each other, but (as predicted by Figure
\ref{fig:colorfixcompare}) the model with a PAH-depleted grain size
distribution in high-density regions (Model B) leads to larger inferred
$\aeff$ for a given $\ks$-4.5$\mu$m color than in Model A, resulting in a
brighter corrected $\mu_{\ks}$.

\begin{figure*}
\plotone{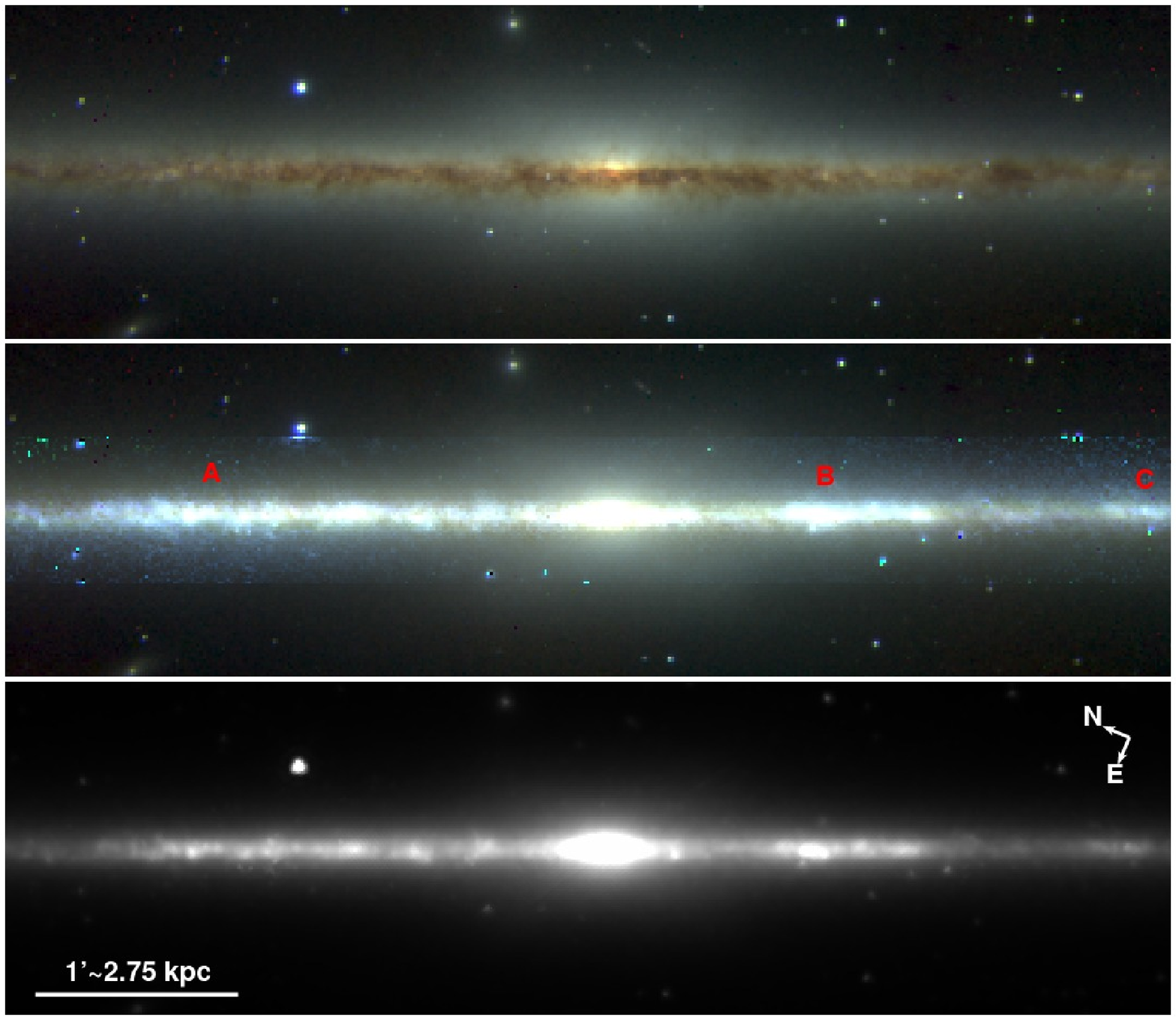}
\caption{Top: central region (5\farcm 8 x 1\farcm 6$\sim$ 16 x 4.4 kpc)of NGC~891. Middle: central region of NGC~891
  after dust correction. Bottom: IRAC 4.5$\mu$m image of the central region of NGC
  891. The top two images are JH$\ks$ color composites with the same
  scaling. Red labels indicate regions outside of the central bulge which
  appear to have excess flux over what would be expected from a smooth
  exponential, possibly indicating star-forming regions or spiral
  structure.}
\label{fig:imagecompare}
\end{figure*}

\section{Results}
\label{sec:sbprofiles}
A false color JH$\ks$ image of the central portion of the dust-corrected galaxy
(using Model B) is shown in Figure
\ref{fig:imagecompare}, compared to the rescaled image before attenuation
correction and the IRAC 4.5$\mu$m image. The central bulge region can now be clearly seen in the
midplane. Additionally, while the majority of the dust-corrected image is
smoothly varying, in several places (marked on Figure \ref{fig:imagecompare})
there appear to be regions of slightly enhanced flux. These regions clearly correspond
to areas with excess flux in the IRAC 4.5$\mu$m image, indicating that they
are real flux overdensities and not artifacts of the attenuation correction
process. These are likely
candidates for star-forming regions. While parts of the three regions are visible
as bright point-source knots of light even before attenuation correction (most
easily visible in Figure \ref{fig:jhkimage}), the true NIR projected size of these
areas is only revealed after attenuation correction. 

In general, the remarkable similarity between the morphology in the
attenuation-corrected $\ks$ image and the IRAC 4.5$\mu$m image is a confirmation
that our correction method is accurate. This similarity is not tautological
since the relation between attenuation and color is based on the RT-model
predictions and not the data.

\subsection{Radial and Vertical Profiles with and without Dust Correction}
\begin{figure}
\plotone{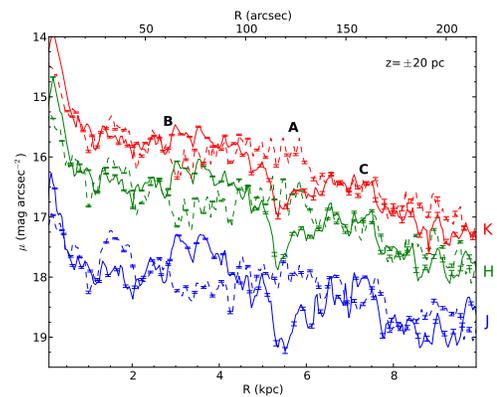}
\caption{Major axis (z=0, vertical aperture size $\sim$40 pc) radial light
  profiles of NGC~891 {\it without} dust correction. Blue, green, and red lines represent J-,
  H-, and $\ks$-band data, respectively. Solid lines show the SW side of NGC~891,
while dashed lines show the NE side. Labels `A', `B', and `C' are at the
positions of the regions of interest shown in Figure \ref{fig:imagecompare}. Error bars are not shown for every point to improve readability.}
\label{fig:radialSBwdust}
\end{figure}

\begin{figure}
\plotone{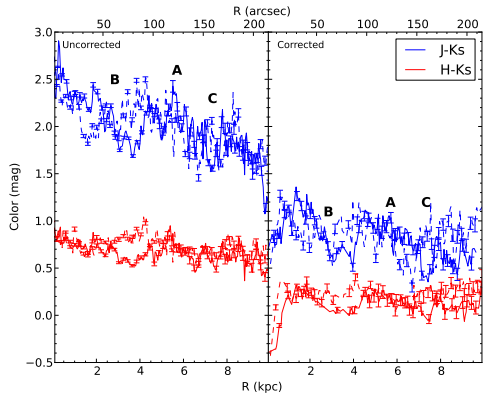}
\caption{Major axis (z = 0, vertical aperture size $\sim$40 pc) J-$\ks$ and
  H-$\ks$ radial color profiles of NGC~891. Line styles, labels, and
  errors are
  the same as for Figure \ref{fig:radialSBwdust}. Left panel shows profiles
  without dust correction, while the right panel shows the profiles after the
  effects of attenuation have been removed.}
\label{fig:radialcolorwdust}
\end{figure}

We show NGC~891's JH$\ks$ radial surface brightness profile along the major axis
(i.e. z=0) un-corrected for dust attenuation in Figure \ref{fig:radialSBwdust}. The vertical bin size is 0.9\arcsec or
$\sim$40 pc. The radial profiles with dust are very similar to those shown in
Figure 9 of
\citet{Aoki91}, except that our data shows a distinct
bump in all bands at $\sim$3 kpc on the SW side while they only see that feature in H and
K. The reason for this discrepancy is unclear, but it indicates that this 3
kpc feature is a stellar over-density, not a dusty spiral arm as hypothesized
by \citet{Aoki91}. We also present un-corrected J-$\ks$ and H-$\ks$ major axis color
profiles in the left panel of Figure
\ref{fig:radialcolorwdust}. The J-$\ks$ profile seems to show a weak trend towards bluer
colors with increasing radius, but there is significant scatter. The
H-$\ks$ profile shows a similarly weak trend. No color trend is visible as a
function of radius when using attenuation corrected profiles (right panel of
Figure \ref{fig:radialcolorwdust}). There is also an
asymmetry between the NE and SW colors of NGC~891 around R$\sim$3 kpc, in
the same position with the apparent stellar overdensity discussed previously
in Figure \ref{fig:radialSBwdust}.

Our major axis radial profiles also show the northeast side of NGC~891, which
was not shown in \citet{Aoki91} due to noise issues in the H band on their
detector. Comparing the radial profiles on both sides of the bulge shows the
significant amount of stochastic structure in the midplane. In addition to the
bumps noted by \citet{Aoki91} the southwest side has another light enhancement
at  $\sim$7.5 kpc and a marked dearth of light at $\sim$5.5 kpc. The northwest
side appears to have fewer of these strong features, however, although it
certainly has its fair share of smaller bumps and wiggles. All of these
features point to the strong observable effect of clumpy dust and starlight,
even at these long wavelengths.

\begin{figure*}
\plotone{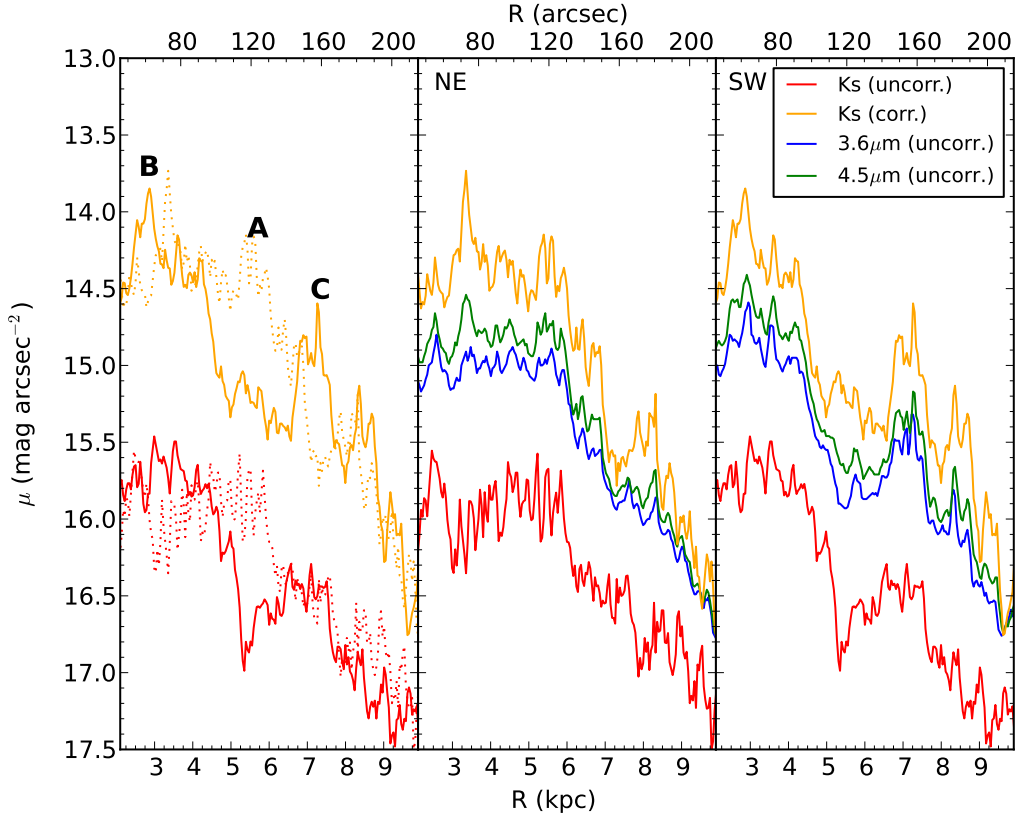}
\caption{Major axis (z = 0, vertical aperture size $\sim$40 pc) radial
  surface-brightness profiles. Left: comparison of profiles before (red) and after (orange) dust
  correction in the $\ks$-band. Solid lines show the SW profile, while dotted
  lines denote the NE side of NGC~891. Labels denote regions of interest discussed in the
  text. Middle (right) panels show uncorrected and corrected $\ks$-band major
  axis profiles compared to IRAC 3.6$\mu$m and 4.5$\mu$m data on the NE (SW)
  sides of NGC~891.}
\label{fig:radialSBdustcompare}
\end{figure*}

We compare the $\ks$-band major axis radial profile from Figure
\ref{fig:radialSBwdust} to its dust-corrected counterpart (using Model B) in
the left panel of 
Figure \ref{fig:radialSBdustcompare}. The dust-corrected profile is
uniformly brighter than the fiducial profile, which is to be expected. Several
local surface-brightness peaks seen in the observed profile are not present in the dust-corrected
profile (at $R\sim 3$ kpc, for instance, on the SW side) while others have
been enhanced. The apparent star-forming regions mentioned in the previous
section all correspond to local maxima in the dust-corrected profile; of
particular note is point `B', which does not stand out on the original
profile. Also interesting is the `trough' on the SW side of the dust-corrected
profile between 4 and 6.5 kpc. While the original profile has a local minima in
the same region, it is much smaller in radial width. It seems likely that striking
appearance of this 
feature is due to a combination of a real intensity drop-off at $R\sim 4$
kpc and the enhancement of the underlying smooth profile at $\sim$7 kpc (point
`C').  

\begin{figure*}
\plotone{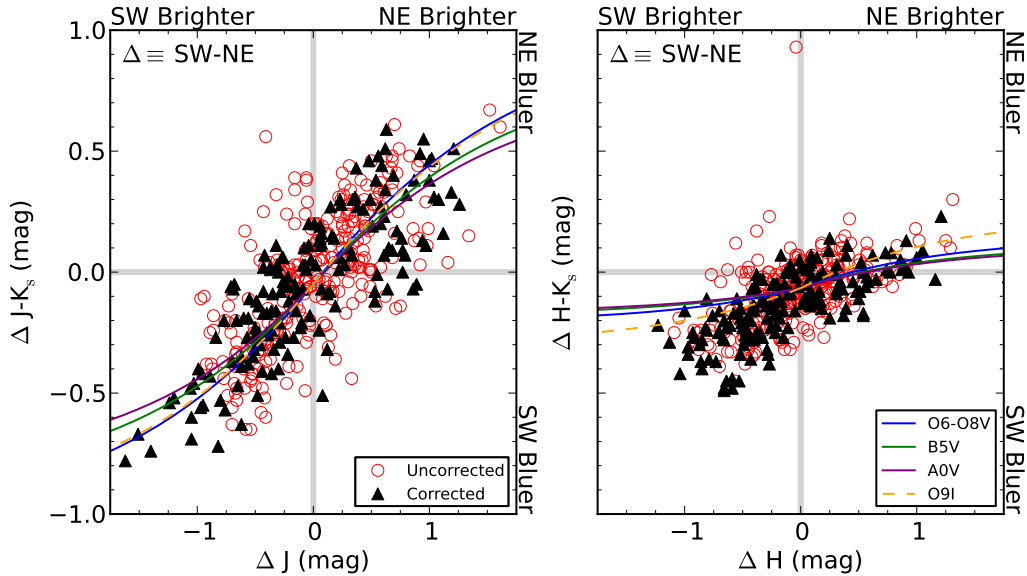}
\caption{Left: J-$\ks$ color differentials between SW and NE sides of NGC~891's major
  axis as a function of J magnitude differentials. Right: same, but for
  H-$\ks$ as a function of H. Red open circles and black filled triangles
  denote data before and after dust correction, respectively. Lines show how
  the colors would change if stars of a given spectral type (from
  \citealt{Tokunaga00}) are added to
  either the SW or NE side of the galaxy. The J-$\ks$ color variance is
  well explained by a late O-type star, although none of the stellar templates
are able to fully match the H-$\ks$ data.}
\label{fig:majoraxisdeltas}
\end{figure*}

Just from looking at the attenuation-corrected radial surface-brightness and
color profiles (Figures \ref{fig:radialcolorwdust} and
\ref{fig:radialSBdustcompare}) it is difficult to discern whether the asymmetries
in the flux and color profiles are due to stellar density enhancements,
intrinsic color variations, or
both. Figure \ref{fig:majoraxisdeltas} shows the magnitude and color
differentials between the SW and NE sides of the major axis profiles, where
$\Delta X$ is defined as the magnitude difference between corresponding
$\sim$40 pc$^{2}$ cells at the same radii but on opposite sides of NGC~891's midplane.
Figure \ref{fig:majoraxisdeltas} clearly indicates that asymmetries in surface brightness are correlated with
color --- brighter regions of NGC~891's midplane are also bluer. The mean
attenuation-corrected J-$\ks$ and H-$\ks$ colors for $\Delta$J and $\Delta$H = 0 are $\sim$0.77 and
$\sim$0.12 mag, respectively, roughly the color of a K2III star. There are minor ($\sim$0.03 mag) differences in these colors
between the two sides, but are well within the dispersions of the data. 

To approximate the effect of star formation on these $\Delta$ metrics we also
compute how the colors would change if the excess midplane light between the
two sides of NGC~891 is due to a hot star. We find that the $\Delta$J-$\ks$
index is well fit by an O-type star, while the trend for $\Delta$H-$\ks$
vs. $\Delta$H is underpredicted for any stellar type. A full investigation of this minor discrepancy is beyond the scope of this work, but
the general trend is clear. Rather than merely being overdensities of old
stars, asymmetries in the radial major-axis profile of NGC~891 are being
caused by the addition of young stellar populations to the underlying stellar density. In some cases these asymmetries are
likely due to individual regions of enhanced star-formation, but the broad nature of several
features are at much larger scales than would be expected for localized
star-forming regions. For instance, the surface brightness
is relatively constant on the NE side of NGC~891 between 3 and 6 kpc, while
the aforementioned SW `trough' is $\sim$2 kpc wide. The simplest explanation
for these features is that we are viewing projected spiral structure, which
NGC~891 is believed to possess \citep{Kamphuis07}.

\begin{figure}
\plotone{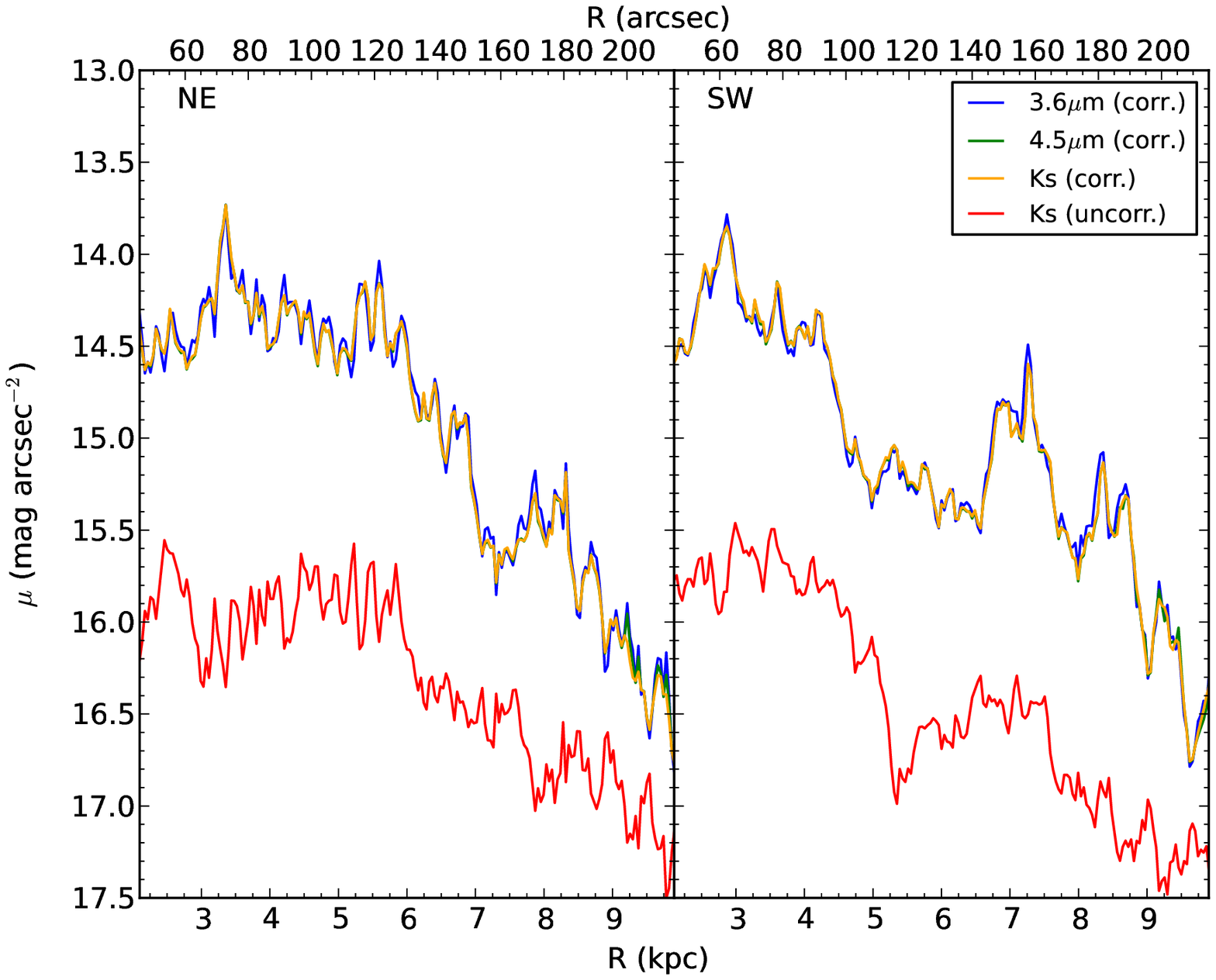}
\caption{Same as the middle and right panels of Figure
  \ref{fig:radialSBdustcompare}, but with attenuation corrected IRAC 3.6 and
  4.5$\mu$m profiles.}
\label{fig:radialSBdustcompare_iraccorr}
\end{figure}

To verify the validity of our dust correction we compare IRAC 3.6 and
4.5$\mu$m major axis profiles to the dust-corrected $\ks$ profile in the middle
and right panels of Figure \ref{fig:radialSBdustcompare}. The IRAC data in
general shows the same non-axisymmetric features as the dust-corrected $\ks$-band
profile, with a few local exceptions (e.g. at $\sim$3 and $\sim$5.5
kpc on the SW side). The IRAC profiles are also globally fainter by up to 0.5 mag
arcsec$^{-2}$, whereas for starlight the unattenuated $\ks$-3.6/4.5$\mu$ color
should be zero (in our Vega photometric system) due to all filters being in the Rayleigh-Jeans regime. These discrepancies are {\it entirely} due to dust
attenuation in the 3.6 and 4.5$\mu$m bands: we computed dust corrections for
these filters in the same way as for the NIR, which when applied to the IRAC major
axis profiles produces profiles virtually identical to the $\ks$-band data
(Figure \ref{fig:radialSBdustcompare_iraccorr}). 

\begin{figure}
\plotone{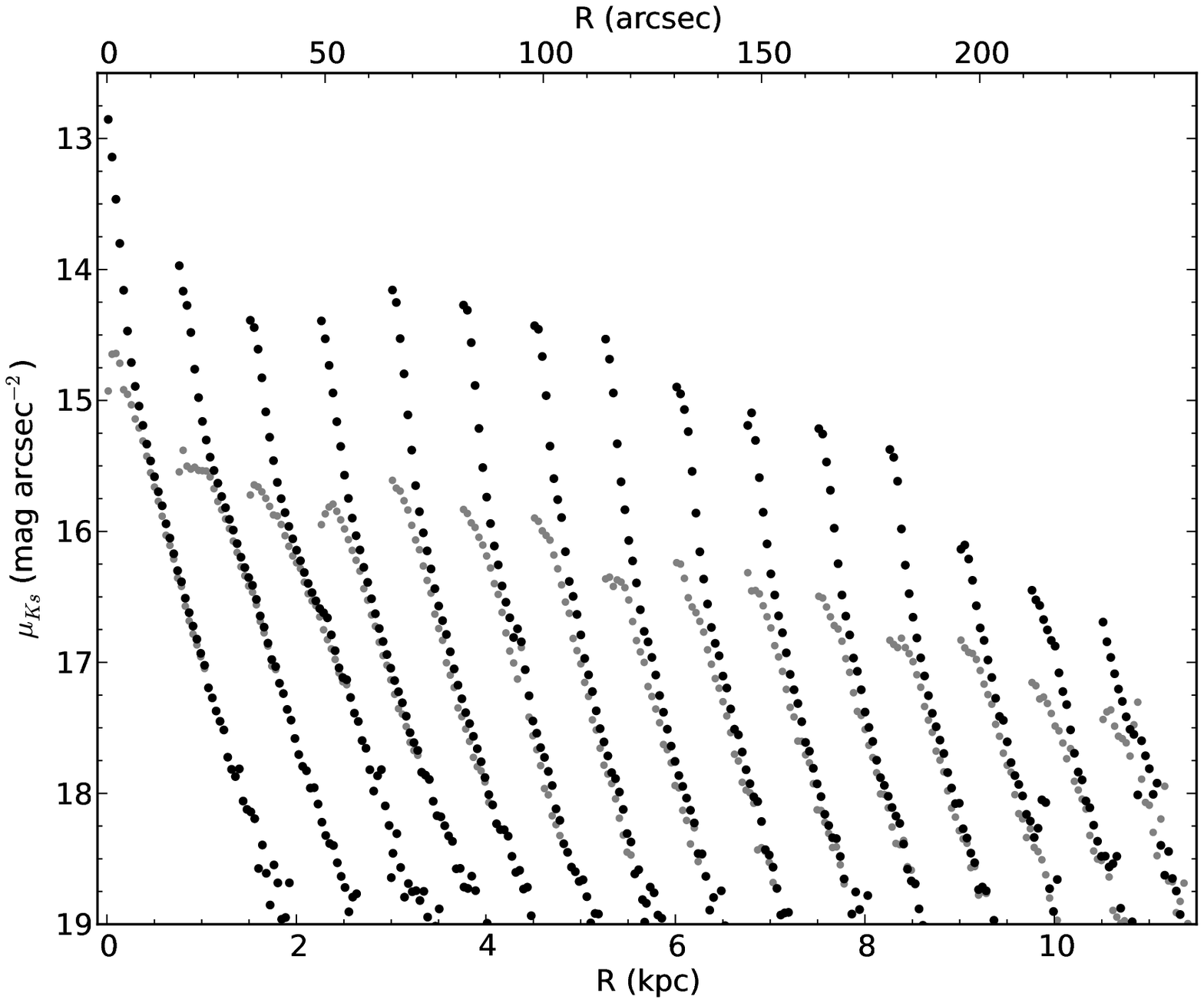}
\caption{Vertical $\ks$-band profiles of NGC~891. All four quadrants of the
  galaxy have been averaged together. The leftmost point of each
  profile is at the radius of that profile, starting at R = 0 kpc and stepped
  every 0.75 kpc. Gray points are the uncorrected
  profiles, while the black points have been attenuation corrected. NGC~891
  appears to have a roughly constant midplane attenuation of $\sim$1.5 mag
  from just outside the very center of the galaxy to 9 kpc in radius. At radii larger than 9 kpc, both the intrinsic
  amount of
light and dust attenuation fall off at a much steeper rate. }
\label{fig:vanderkruitplot}
\end{figure}

A series of $\ks$-band vertical profiles as several radii are shown in Figure
\ref{fig:vanderkruitplot}. Gray points show the profiles including the effects
of dust, while black points have been attenuation corrected. The midplane
attenuation is roughly constant at $\sim$1.5 mag arcsec$^{-2}$ out to R = 8.5 kpc. The amount of attenuation quickly decreases with z, however, and
by z$\sim$0.5 kpc the effects of dust on the $\ks$-band vertical profiles are
almost negligible. At R$\sim$9 kpc there is a truncation in the central SB of
the dust corrected profiles but not of the uncorrected ones. This scenario can
only result from a cutoff in {\it both} the light near the midplane as well as
the dust attenuation. This feature illustrates the importance of properly
accounting for dust attenuation, as the uncorrected profile appears to be
smoothly decaying across this threshold. 

\subsection{Image Fitting}

\subsubsection{Model Profiles}
\label{sec:profiles}
The intensity distribution of an edge-on spiral with a radially
exponential disk luminosity density distribution at all heights can be
described, in the absence of attenuation, by
the analytic function
\begin{equation}
I(R,z) = I_{0}\frac{|R|}{h_{R}}K_{1}\left( \frac{|R|}{h_{R}}\right) f(z),
\label{eq:sbdistro}
\end{equation}
where $I_{0}$ is the central flux, $K_{1}$ is the modified Bessel function of
the first order, and $f(z)$ is a function governing the vertical light
distribution \citep{vanderKruit81a}.

A simple analytic description of the vertical density profile of a
spiral galaxy has been proposed by \citet{vanderKruit88} to be of the form:
\begin{equation}
f(z) = 2^{-2/n}\mathrm{sech}^{2/n}\left(n|z|/(2h_{z})\right),
\label{eq:vertprof}
\end{equation}
where $n$ varies between 1 and $\infty$. The limiting behavior of this formula
produces a sech$^2$ distribution
\begin{equation}
f(z) = \frac{\mathrm{sech}^{2}(|z|/(2h_{z}))}{4}
\end{equation}
when $n$=1 and the exponential distribution
\begin{equation}
f(z) = \mathrm{exp}(-|z|/h_z)
\label{eq:exponential}
\end{equation}
when $n$ = $\infty$. For convenience we include the factor of 4 in the
sech$^{2}$ profile with $I_{0}$ and use 
\begin{equation}
f(z) = \mathrm{sech}^{2}(|z|/(2h_{z}))
\label{eq:isotherm}
\end{equation}
instead.

The first attempts to fit the detailed vertical surface-brightness profiles of
edge-on spirals began by treating the disk as a locally isothermal
self-gravitating sheet \citep{vanderKruit81a},
which results naturally in the sech$^2$ model \citep{Spitzer42}. As a result this model is
generally referred to as the `isothermal' distribution. Note, however, that
this strict physical connection is lost when considering multiple disks,
rotating disks with radial gradients, and/or disks embedded in a dark matter
halo. Nonetheless, the sech$^{2}$ profile is not a bad descriptor of many
observed disk vertical light profiles.
More recent work has found that a steeper profile appears to fit some
data better --- usually an
exponential profile \citep{Wainscoat89,Aoki91,Xilouris99}, although sometimes the intermediate $n=2$ `sech' profile
is employed \citep{Barnaby92,Rice96}. A theoretical
framework which produces vertical surface-brightness
distribution much closer to that of an exponential from physical conditions in
spiral galaxies (by considering the gas in addition to the stars) has been developed
partly in direct response to these experimental findings
\citep{Banerjee07}. \citet{Comeron11b} also find that the vertical density
profiles for dynamically consistent models of multiple self-gravitating disks
are steeper than a sech$^{2}$ function. Discussed further in Section
\ref{sec:expvssech2}, their models are well characterized by the intermediate sech vertical profile.

In general, however, there is currently no concrete consensus for which type of
profile (or superposition of profiles) better represents the underlying
distribution of stars, due largely to the fact that both the exponential and
sech$^{2}$ profiles have significant deviations from each other only at small
heights, exactly the place where dust obscuration is most prevalent. In this
work, since we are able to accurately remove the deleterious effects of dust on the data,
we will use both exponential and sech$^{2}$ profiles in our vertical fits and
see if we are able to distinguish one as clearly better in a statistical
sense. Additionally, since NGC~891's thin and thick disks are well known
(e.g. \citealt{Morrison97,Ibata09}) and there is a clear super-exponentiality
in the dust corrected profiles near the midplane in Figures \ref{fig:correctioncompare},
\ref{fig:imagecompare}, and \ref{fig:vanderkruitplot}, we
expect our fits to require three separate vertical components in order to
constrain the data at all z. 

\subsubsection{Seeing Convolution}
Normally, observational studies of edge-on spirals have not required a detailed treatment of
the effects of seeing on the observed profiles. This is a result of the
combination of large pixel-sizes on detectors and the censoring of regions
near the midplane to avoid dust effects
(e.g. \citealt{deGrijs96};\citealt{deGrijs97}). However, in this work the
seeing correction is of significant importance, especially in distinguishing
between sech$^{2}$ and exponential fits, where the seeing serves to smooth
out the exponential profile and make it appear more like a sech$^{2}$.

We make the simplifying assumption that the seeing profile can be approximated
by a 2D Gaussian point-spread-function, as in \cite{Pritchet81}. Then, the
convolution is
\begin{align}
I_{conv}(R,z) = 
&\frac{1}{2\pi\sigma^2}\int_{-\infty}^{\infty}I(r',z')\, \mathrm{exp}\left( \frac{-(R-r')^{2}}{2\sigma^{2}}\right) \nonumber \\
&\, \mathrm{exp}\left(\frac{-(z-z')^{2}}{2\sigma^{2}}\right)\,dr'dz',
\end{align}
where $\sigma$ is the FWHM of the seeing. Since all the profiles we are considering here are of the form of equation
\ref{eq:sbdistro} we can perform the convolution
separately in $R$ and $z$:
\begin{align}
I_{conv}(R,z) =
&\frac{I_{0}}{2\pi\sigma^2}\int_{-\infty}^{\infty}\frac{|r'|}{h_{R}}K_{1}\left(
  \frac{|r'|}{h_{R}}\right)\,\mathrm{exp}\left(
  \frac{-(R-r')^{2}}{2\sigma^{2}}\right)\,dr' \nonumber \\
&\times \int_{-\infty}^{\infty}f(z)\,\mathrm{exp}\left(\frac{-(z-z')^{2}}{2\sigma^{2}}\right)\,dz'.
\label{eq:splitconv}
\end{align}
 For both exponential and sech$^{2}$ profiles
we used the `convolve' function in the numpy Python library. For every value
to be convolved the surface-brightness profile was sampled with a spacing of
0.1 times the seeing (FWHM = 0\farcs 9) over a range of values equal to six times the seeing in
either direction. The Gaussian kernel was sampled at the same spacing between -3 and
+3 times the seeing. This scheme ensured that the convolution did not miss any
significant amounts of flux due to edge effects from either the convolution
kernel or the function, while minimizing the computation time required to
perform the integral.

When the relevant scale-length or -height is much larger than the seeing, the effect of the convolution on the
original profile is
negligible. This is especially true at large distances from the center of the
profile. Since the radial scale-length is much larger than the seeing we use the unconvolved radial intensity profile in our fitting analyses to
minimize computation time.

\subsubsection{2D Fitting}
\label{sec:2dfitting}
With the surface-brightness profiles corrected for the attenuation and
convolutions for seeing in place, we now
fit the intrinsic distribution of starlight. While one-dimensional fits have
been successfully employed to compare single-component surface-brightness
models to data, \citet{Morrison97} make a compelling argument, based on
degeneracies between free parameters, for the
necessity of fully 2D fits when multiple stellar distributions overlap. We
therefore opt to model our data exclusively with image-based fits. We choose
to model the $\ks$-band, as it is where both the dust correction and stellar
population-mismatch errors will be smallest (due to the decreasing attenuation
with increasing wavelength and all stellar fluxes being on the Rayleigh-Jeans tail).

Before performing our fits, however, we first mask the image based on several
criteria. All pixels with flux
from foreground or background contaminants are masked, as well as pixels with
errors $>$ 1 mag (while large-error pixels can still contribute to a
$\chi^{2}$ fit in aggregate, gains from leaving these pixels in will be
minimal and removing them allows for computational speedup in the fitting process). Contaminant masking was done by
hand and whenever possible we choose to be conservative. We also mask out all
pixels with $R < 3$ kpc to remove any pixels dominated by ``bulge'' light. We
refer to the bulge light in quotations because, as we shall see in Section
\ref{sec:fitcentral}, the central light distribution is complex and likely
non-axisymmetric. For exactly these reasons we wish to focus first on the
outer (in radius) disks. 

We use the Python lmfit\footnote{http://newville.github.com/lmfit-py/} extension to
the Levenberg-Marquardt nonlinear least-squares minimization routine in
scipy.optimize.leastsq (which is itself based on the MINPACK-1
library). Although we report surface-brightness values in mag arcsec$^{-2}$, all
fitting is performed on flux values to preserve
error estimates. To avoid false minima, we ran the fit 100 times for 1- and 2-component models and 200 times for 3-component models. The initial guesses of all free parameters for each run were chosen
randomly within upper and lower limits based on what we considered to be
reasonable ranges for spiral galaxies (Table \ref{tab:2dfit_newdust}). The model with
the lowest $\chi^{2}$ was selected and used as the input guess to 100
bootstrapping iterations, finding, as \citet{Morrison94} did, that standard
error estimates from the nonlinear fit were often too small. Note that only
random uncertainties are shown in Table \ref{tab:2dfit_newdust}; the
$\lesssim$8\% systematic error on the central surface brightnesses from input
stellar SED mismatch (Section \ref{sec:multidust}) is not shown.

\begin{deluxetable*}{ccccccccccccc}
\tablewidth{0pt}
\tabcolsep 1.8pt
\tablecaption{Two-Dimensional Fits}
\tablehead{ \colhead{} & \multicolumn{2}{c}{Allowed Range} & \colhead{} &
  \multicolumn{4}{c}{One Disk} & \colhead{} & \multicolumn{4}{c}{Two Disks}
\\
\cline{2-3} \cline{5-8} \cline{10-13}
 \colhead{Parameter$^{\mathrm{a}}$} &\colhead{Min} & \colhead{Max} & \colhead{} & \colhead{exp}
& \colhead{$\Delta^{\mathrm{b}}$} & \colhead{sech$^{2}$} & \colhead{$\Delta^{\mathrm{b}}$} & \colhead{} & \colhead{exp} & \colhead{$\Delta^{\mathrm{b}}$} & \colhead{sech$^{2}$} & \colhead{$\Delta^{\mathrm{b}}$}} 
\startdata
$\mu_{0,1}$&25&10&&14.00$\pm$0.018&0.04&14.52$\pm$0.020&0.13&&13.26$\pm$0.040&-0.58&13.53$\pm$0.026&-0.66\\
$h_{R,1}$&0.1&10&&3.55$\pm$0.027&0.12&3.67$\pm$0.025&0.11&&2.42$\pm$0.045&-0.89&2.54$\pm$0.032&-0.70\\
$h_{z,1}$&0.01&2.5&&0.28$\pm$0.002&0.00&0.19$\pm$0.002&0.01&&0.12$\pm$0.004&-0.10&0.07$\pm$0.001&-0.02\\
$L_{tot,1}^{\mathrm{c}}$&\nodata&\nodata&&1.37$\times 10^{11}$(100)&\nodata&1.19$\times 10^{11}$(100)&\nodata&&7.92$\times 10^{10}$(47)&\nodata&7.57$\times 10^{10}$(48)&\nodata\\
$\mu_{0,2}$&25&10&&\nodata&\nodata&\nodata&\nodata&&15.29$\pm$0.101&-1.91&15.68$\pm$0.067&0.07\\
$h_{R,2}$&0.1&10&&\nodata&\nodata&\nodata&\nodata&&4.44$\pm$0.241&0.63&4.30$\pm$0.178&0.89\\
$h_{z,2}$&0.01&2.5&&\nodata&\nodata&\nodata&\nodata&&0.48$\pm$0.016&-0.77&0.33$\pm$0.007&-0.02\\
$L_{tot,2}^{\mathrm{c}}$&\nodata&\nodata&&\nodata&\nodata&\nodata&\nodata&&8.97$\times 10^{10}$(53)&\nodata&8.33$\times 10^{10}$(52)&\nodata\\
$\Delta$z&$-$0.1&0.1&&0.05$\pm$0.001&0.00&0.05$\pm$0.001&0.00&&0.05$\pm$0.001&0.00&0.05$\pm$0.001&0.00\\
Reduced $\chi^{2}$&\nodata&\nodata&&7.32&1.09&11.7&1.70&&4.21&-0.19&4.61&-0.5\\
\enddata
\tablenotetext{a}{$\mu$ values in mag arcsec$^{-1}$, $h_{R}$, $h_z$, and $\Delta$z values in kpc.}
\tablenotetext{b}{Reported values in `exp' and `sech2' columns are for Model B,
  and $\Delta=\mathrm{Parameter}_{\mathrm{Model\, B}} -
  \mathrm{Parameter}_{\mathrm{Model\, A}}$.}
\tablenotetext{c}{Total luminosities are given in solar units where $L_{\odot,K}=3.909\times 10^{11} W Hz^{-1}$ (from
  \citealt{Binney98}). Values in parenthesis are the percentage of the total
  disk luminosity in that component.}
\tablenotetext{d}{Fixed values from \citet{Ibata09}.}
\label{tab:2dfit_newdust}
\end{deluxetable*}

\addtocounter{table}{-1}

\begin{deluxetable*}{cccccccc}
\tablewidth{0pt}
\tabcolsep 1.8pt
\tablecaption{Two-Dimensional Fits, Continued}
\tablehead{ \colhead{} & \multicolumn{2}{c}{Allowed Range} & \colhead{} & \multicolumn{4}{c}{Three Disks} 
\\
\cline{2-3} \cline{5-8}
 \colhead{Parameter$^{\mathrm{a}}$} &\colhead{Min} & \colhead{Max} & \colhead{} & \colhead{exp}
& \colhead{$\Delta^{\mathrm{b}}$} & \colhead{sech$^{2}$} & \colhead{$\Delta^{\mathrm{b}}$} }
\startdata
$\mu_{0,1}$&25&10&&13.08$\pm$0.040&-0.94&13.41$\pm$0.023&-0.67\\
$h_{R,1}$&0.1&10&&1.96$\pm$0.038&0.13&2.32$\pm$0.022&-0.37\\
$h_{z,1}$&0.01&2.5&&0.08$\pm$0.004&0.02&0.06$\pm$0.001&0.00\\
$L_{tot,1}^{\mathrm{c}}$&\nodata&\nodata&&5.05$\times 10^{10}$(29)&\nodata&6.62$\times 10^{10}$(38)&\nodata\\
$\mu_{0,2}$&25&10&&14.52$\pm$0.096&0.47&15.32$\pm$0.067&0.35\\
$h_{R,2}$&0.1&10&&3.87$\pm$0.118&0.46&4.11$\pm$0.114&0.51\\
$h_{z,2}$&0.01&2.5&&0.29$\pm$0.010&0.04&0.25$\pm$0.006&0.03\\
$L_{tot,2}^{\mathrm{c}}$&\nodata&\nodata&&9.60$\times 10^{10}$(55)&\nodata&8.41$\times 10^{10}$(48)&\nodata\\
$\mu_{0,3}$&25&10&&17.81$\pm$0.015&0.04&18.70$\pm$0.015&0.05\\
$h_{R,3}$&\nodata&\nodata&&4.80$^{\mathrm{d}}$&\nodata&4.80$^{\mathrm{d}}$&\nodata\\
$h_{z,3}$&\nodata&\nodata&&1.44$^{\mathrm{d}}$&\nodata&1.44$^{\mathrm{d}}$&\nodata\\
$L_{tot,3}^{\mathrm{c}}$&\nodata&\nodata&&2.85$\times 10^{10}$(16)&\nodata&2.52$\times 10^{10}$(14)&\nodata\\
$\Delta$z&$-$0.1&0.1&&0.05$\pm$0.001&0.00&0.05$\pm$0.001&0.00\\
Reduced $\chi^{2}$&\nodata&\nodata&&3.77&-0.47&3.65&-0.55\\
\enddata
\label{tab:2dfit_newdust_cont}
\end{deluxetable*}

We fit models with 1-, 2-, and 3-components for both attenuation corrections,
motivated by the presence of three disk components in the MW: the young,
star-forming disk, the thin disk, and the thick disk (\citealt{vanDokkum94},
although \citealt{Bovy12} indicate that perhaps the boundaries between the disks are not so
well-defined). It is important to note that we do not force the physical
parameters of the super-thin or thin disks to resemble those seen in NGC~891 and/or the MW by
prior studies by way of limiting our parameter space (although we do restrict
the parameter space for the thick disk; see next paragraph for details). On the contrary our
externally imposed constraints are extremely permissive and allow for fitted
values that can either agree or disagree with the literature.

For models with one or two
components each component had three
free parameters: $\mu_{0}$, $h_{R}$, and $h_{z}$. Additionally, a single
vertical offset for each model was fit simultaneously, to remove the effects
of any small errors from our choice of galaxy center. Because any such errors
in R will be much smaller than $h_{R}$ (and therefore have minimal effect on
the fits) to avoid additional computational overhead we do not allow a radial
offset in the fits. Therefore, for a model with 1 (2) vertical component(s),
there are 4 (7) free parameters. Preliminary 3-component
fits were unable to consistently select physically plausible fits due to the
low S/N at large heights; however,
visual inspection of the 2-component residuals showed an excess of light
at large heights above the midplane, especially when fitting with sech$^{2}$
profiles. This arises from the shallower inner profile of the sech$^{2}$
function which drives the fits (in a $\chi^{2}$ sense) to a smaller $h_{z}$. To reduce our parameter space (and therefore the
degeneracies) we restrict the scale-length and -height of the 3rd (most extended) component to values given by \citet{Ibata09}, who used HST to obtain very precise
scale parameters with resolved star-counts of NGC~891's thick disk. We do,
however, leave the flux normalization for this component a free parameter, to
allow for band-pass differences and an incomplete understanding of the
integrated SED from the star-count analysis. Therefore our 3-component fits
involve 8 free parameters. The results from all
fits are shown in Table \ref{tab:2dfit_newdust}.

Generally there is very good agreement between the fits from the two
attenuation correction models. The 1-component fits are nearly identical,
especially for the exponential case. The main distinction of note for the 2
and 3-component models is that the central surface-brightnesses of the
innermost (super-thin) disk fits are smaller for Model A than for Model B, which is to be
expected given the increased amount of attenuation correction for a given
color in Model B (see Figures \ref{fig:colorfixcompare} and
\ref{fig:correctioncompare}). Since none of the differences in fitted
parameters between Model A and B would substantially affect our initial
analysis (although the differences between the models are more relevant for
our discussion of the total disk luminosities; see Section \ref{sec:totlum}), for
simplicity we will focus the discussion on the fits from Model B because it
appears to provide a better fit to the colors of the data (c.f. Figures
\ref{fig:cc36profiles}-\ref{fig:cc45profiles} and Section \ref{sec:multidust}).

\begin{figure*}
  \plotone{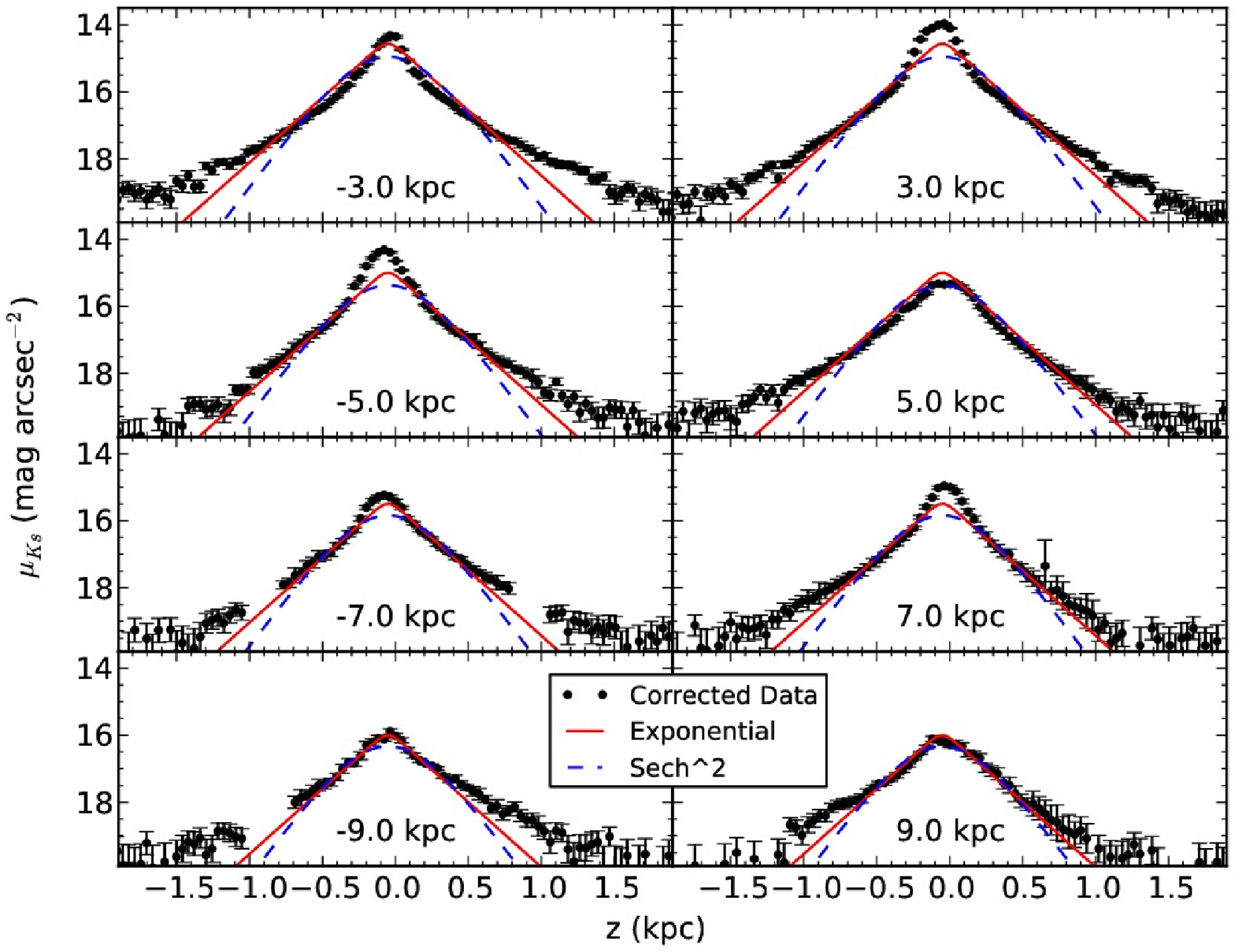}
  \caption{1D vertical slices of both the $\ks$-band dust-corrected data (Model B) and the 2D fit for a
    single disk component. Data and models are shown at four radial locations
    on both halves of NGC~891, with radial bins of $\sim$40 pc. Data are shown as black points, fits are
    denoted by red solid (exponential) and blue dashed (sech$^{2}$)
    lines. Positive radii are on the SW side of NGC~891.}
\label{fig:1compslice}
\end{figure*}

\begin{figure*}
  \plotone{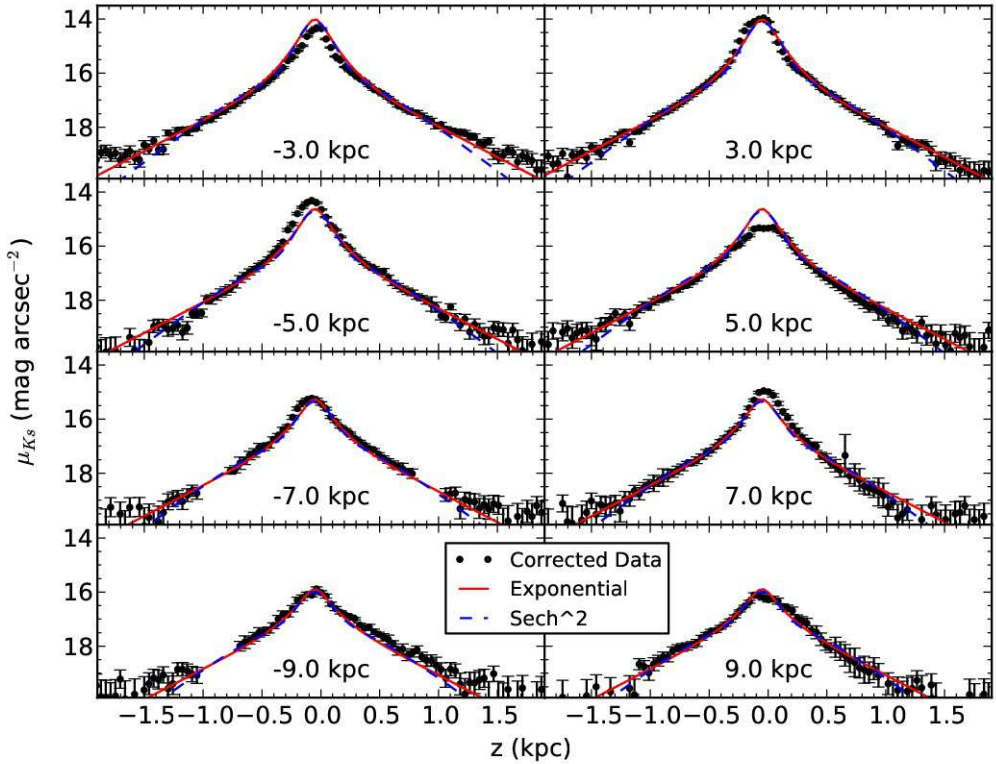}
  \caption{Same as Figure \ref{fig:1compslice}, but for two disk components.}
\label{fig:2compslice}
\end{figure*}

\begin{figure*}
  \plotone{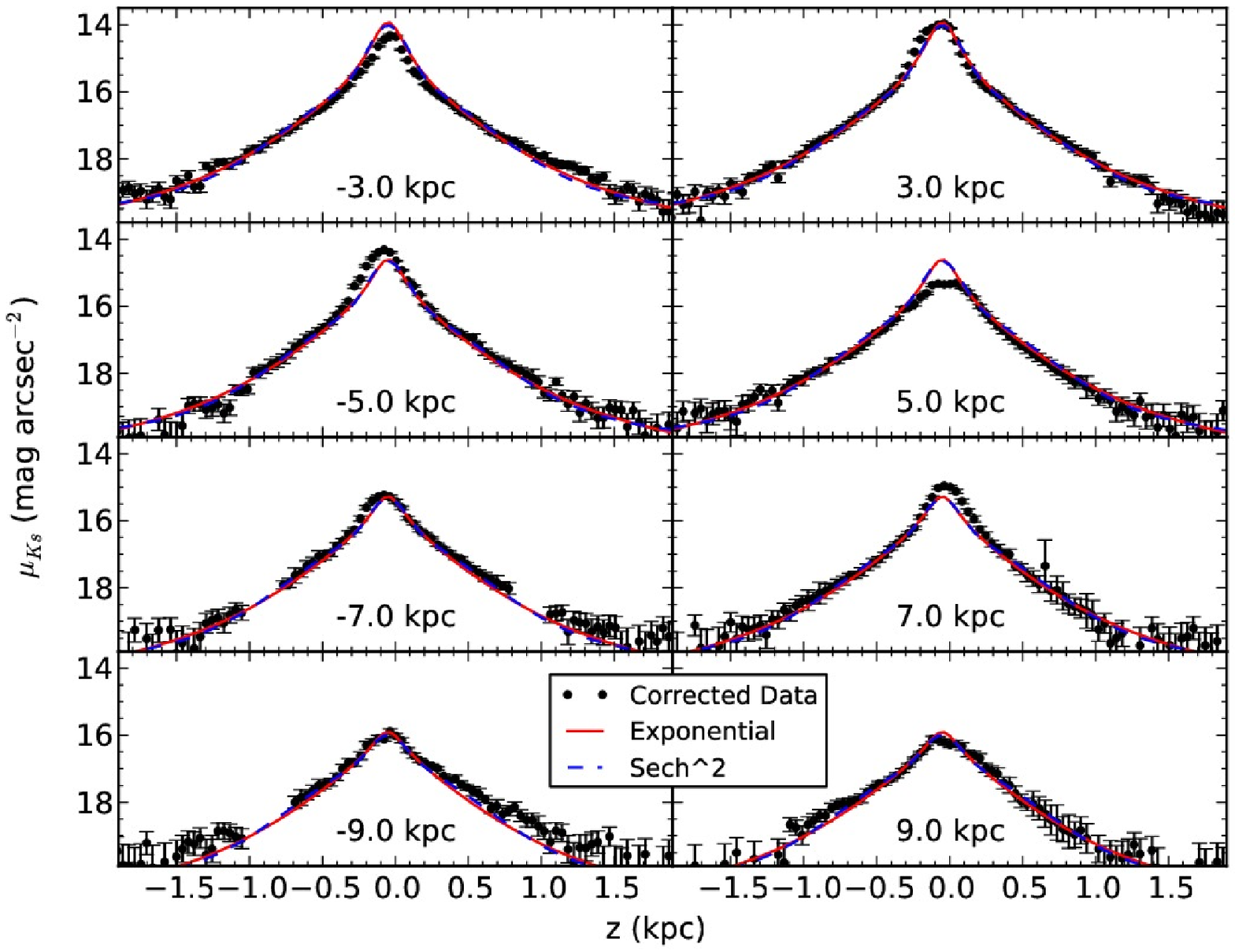}
  \caption{Same as Figure \ref{fig:1compslice}, but for three disk components.}
\label{fig:3compslice}
\end{figure*}

Several vertical slices of the data, overlaid with the relevant 1D portions of
the fits, are shown in Figures \ref{fig:1compslice}-\ref{fig:3compslice}. The
single component fits (Figure \ref{fig:1compslice}) are clearly too simple, failing to match
the peak intensity at the midplane as well as at high latitude. Figure \ref{fig:2compslice} illustrates the dramatic improvement between the 1
and 2-component fits. Whereas the single component fit had difficulty fitting
the data at any height, the 2-component sech$^{2}$ fit does a reasonable job
within $\pm$0.5 kpc, while the exponentials appear to accurately fit almost the
entire profile. The difference in reduced $\chi^{2}$ between using
sech$^{2}$ functions versus exponentials is relatively small compared to the statistical
improvement in adding the second component to the fits, however, since the
regions of largest discrepancy in the 2-component fits are where the noise is
largest. This discrepancy arises because a single exponential (even with the seeing
correction) is more peaked than a sech$^{2}$ and is therefore better able to fit
the (most statistically relevant) midplane regions, leaving the second
exponential to fit the outer regions.

\begin{figure*}
\plotone{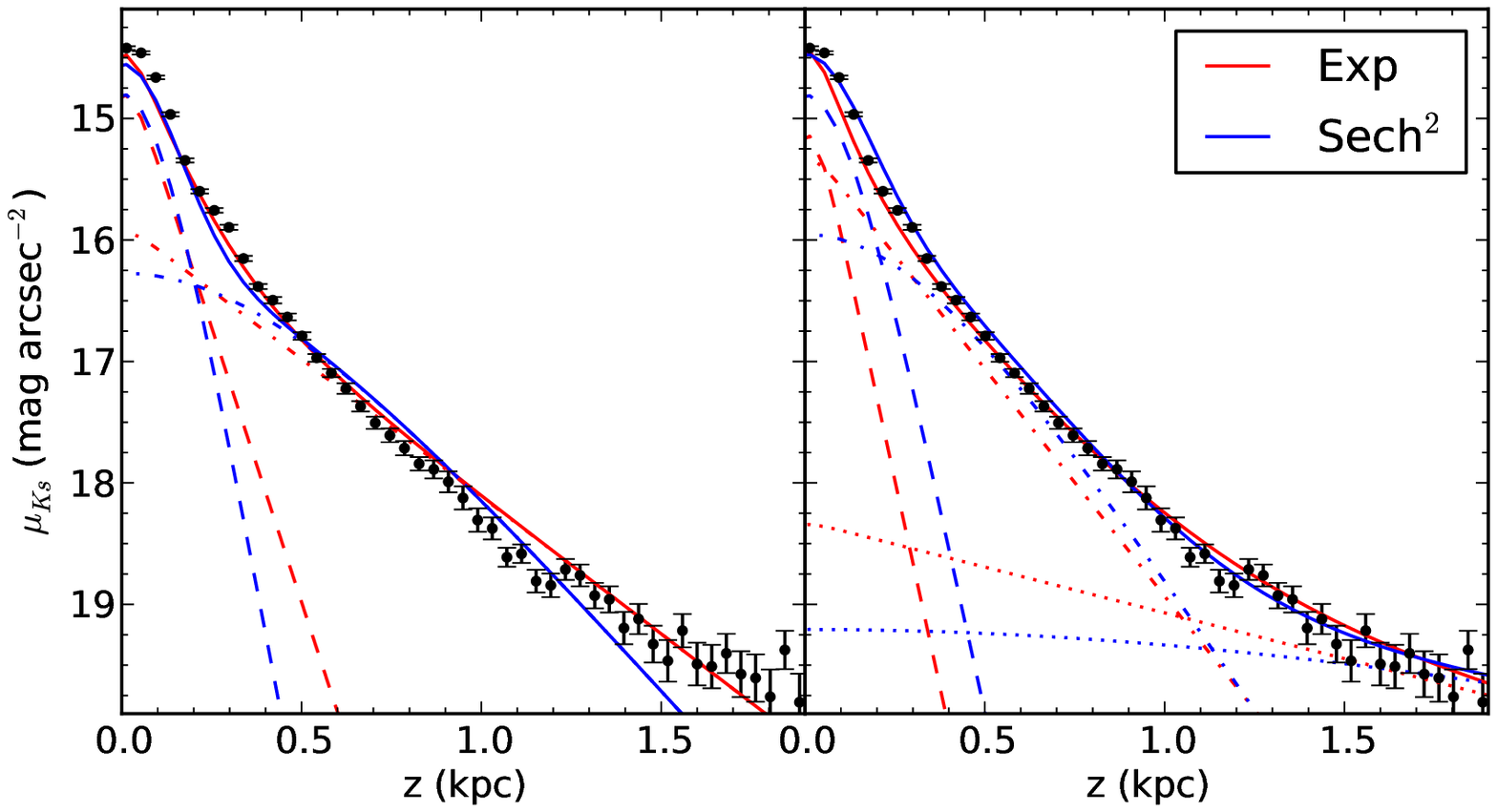}
\caption{Representative vertical profile for NGC~891 (R=4.5$\pm$0.04 kpc), overlaid
  with the best-fitting two-component (left) and three-component (right) seeing-convolved exponential (red) and sech$^{2}$ (blue) models. Black
  points are data (averaged over all four quadrants of the galaxy), while solid lines show the total surface-brightness
  profiles of the models. The surface-brightness profiles of the individual
  disks which make up the models are denoted by dashed, dot-dashed, and dotted
  lines (showing the disks in order of smallest to largest $h_{z}$). The data have
  been averaged over all four quadrants of the galaxy.}
\label{fig:fitdecomp}
\end{figure*}

While adding a third component decreases the reduced $\chi^{2}$ of the
exponential fits only by a small amount, the improvement in the sech$^{2}$
fits is more significant, both in the $\chi^{2}$ sense and upon visual
inspection of Figure \ref{fig:3compslice}. The 3-component exponential and sech$^{2}$ fits are virtually
indistinguishable from each other, both in a visual and statistical
sense, and both are clearly able to fit the vertical profile at all heights. A representative plot of vertical surface brightness fits with the
contributions of the individual disks is shown in Figure
\ref{fig:fitdecomp}. First, the visual improvement resulting from adding the
third component is clearly noticeable, especially at heights above $\sim$0.75
kpc. Additionally, while the total surface-brightness profiles of the 3-component exponential and
sech$^{2}$ fits are very similar, there are noticeable differences in the form
of the individual components. These differences are largest in the relative
$\mu_{0}$ between the exponentials and the sech$^{2}$s, mostly to compensate for
the flatter central profile of the sech$^{2}$ profiles. 

Even with the 3-component models, we find reduced $\chi^{2}>1$. This could be
due to an underestimation of our errors, most likely in the attenuation
correction. However, given the relative simplicity of our model compared to
the complex nature of a galactic system (e.g. the model is perfectly axisymmetric), the fact that we do not recover a
reduced $\chi^{2}\sim 1$ is not surprising. Our 2D fits do not take into
account the potential effect of warps, star-forming regions, spiral arms, or
any other non-axisymmetric structure. Despite these limitations, however, the
reduced $\chi^{2}$ is still effective at distinguishing between the relative
goodness-of-fit of models, as shown by the improvement from 1-component fits to 3-component fits both in reduced $\chi^{2}$ and upon visual inspection.

\subsubsection{Disk Asymmetries}
\begin{figure*}
  \plotone{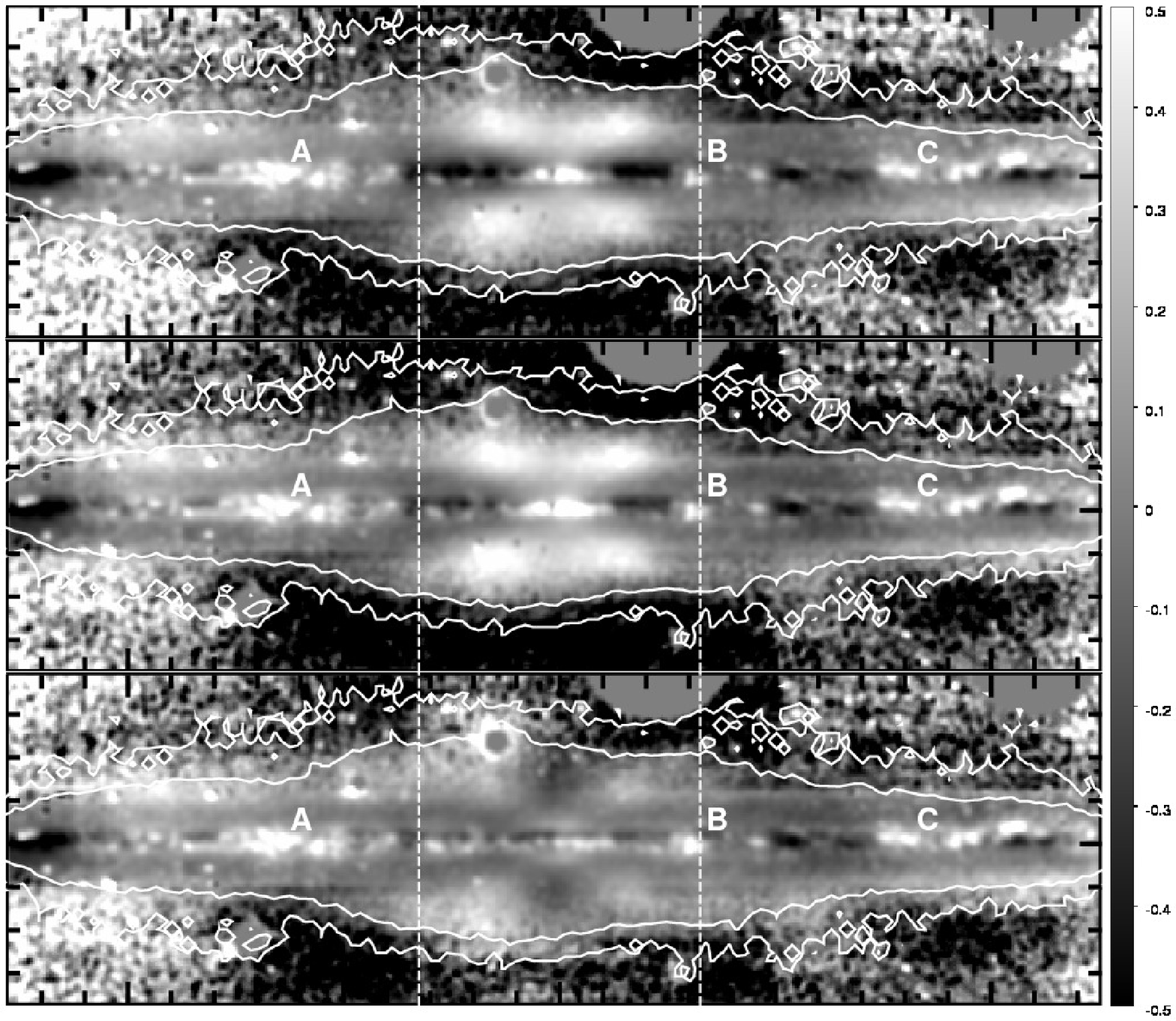}
  \caption{Top two panels: Residual surface brightness image (model - dust-corrected data) for
    the 3-component models in Table \ref{tab:2dfit_newdust} (Model B
    attenuation correction). Top:
    model with exponential vertical light distribution. Middle: model with
    sech$^{2}$ vertical light distribution. The bottom plot shows the residual
    image for the model which adds a nuclear disk, bar, and bar/disk
    truncation to the 3-component model (Table \ref{tab:bulgefits}). The data
    have been masked as they are in the 2D fitting algorithm (except for
    the inner 3 kpc region, not masked here, which is simply denoted by dashed lines), and
    smoothed by a Gaussian with $\sigma$=2\farcs 7. Colorbar
    units are mag arcsec$^{-2}$. White contours denote 0.2 and 0.5 mag
    pixel$^{-1}$ errors in the unsmoothed data. White labels denote regions corresponding to areas
    which, based on a visual inspection of the images, appear to contain embedded
  star clusters. the full image extent is (width x height) 23.8 x 9.6 kpc, where
1'$\sim$2.75 kpc. Each tick corresponds to 20''. The vertical dashed lines show the inner limits of the data
used in the model fitting for the 3-component models. The SW side is on the right.}
\label{fig:modelresid}
\end{figure*}

To investigate the impact of non-axisymmetric structures on our model fits to NGC~891, we constructed
model - (dust-corrected) data residual surface-brightness images for both 3-component models
(top two panels of Figure \ref{fig:modelresid}). These images have been
masked (bad values set to zero) in the same way as for the 2D fitting (except
for the inner 3 kpc, the boundaries of which are marked by dashed lines but
not masked out), and
smoothed by a $\sigma$=2\farcs 7 Gaussian to bring out larger-scale
detail. Additionally, the $\ks$-band bulge determined by \citet{Xilouris99} has
been added to the model image before producing the residual. White contours show the
boundaries of regions with errors $<$0.2 and $<$0.5 mag pixel$^{-1}$
(unsmoothed) --- while we fit all points with errors $<$1 mag pixel$^{-1}$, we
weight each pixel by its error and as a result the fit is most sensitive to
pixels with small uncertainties. Therefore we focus our analysis of the
quality of the fit on the regions of the image with smaller uncertainties. 

Several asymmetries are clearly present in Figure \ref{fig:modelresid}. Based on a visual
inspection of the data before resampling, at least three of them
(labeled in Figure \ref{fig:modelresid} and also
shown in Figures \ref{fig:jhkimage} and \ref{fig:imagecompare}) are
likely to be regions of enhanced local star formation in NGC~891. \citet{Oosterloo07} find a mild HI warp in NGC~891 (position angle =
1.5$^{\circ}$). There are some slight indications of this warp in the data,
chiefly in the excess of light over the model above the midplane at radii
larger than region `C' on the SW side. On extremely close inspection of the
non-dust-corrected mosaics (see Figure \ref{fig:jhkimage}, although it is
difficult to distinguish in that image stretch) it is also possible to discern a similar offset in
the dust, but due to the clumpiness of the dust and the faintness of this
feature we are unable to make a stronger claim about whether or not it is real.


\subsubsection{Fitting the Central Region of NGC~891}
\label{sec:fitcentral}
Beyond the specific asymmetries called out above, it is clear that these models do a
poor job of fitting NGC~891's $\ks$-band light within 3 kpc. In addition to the
large amount of light above and below the midplane from the galaxy's bar that
is not reproduced in our model there
is also clearly excess light in NGC~891's nucleus --- even with the addition
of the \citet{Xilouris99} bulge to the model. In contrast, just
outside of the nucleus the model greatly overpredicts the flux near the midplane. The
simplest explanation for this feature is that at R$\sim$3 kpc there is a
transition between disk and bar dominated regions, such that the 3-component
disk is truncated within this radius and the bar does not extend beyond this
radius. There is ample evidence for such truncations in other galaxies
(e.g. \citealt{Anderson04}). To the extent that bars arise from dynamical
instabilities in the inner disk, then this is not a truncation {\it per se},
but a redistribution of the disk from an axisymmetric, dynamically cold stellar mass
distribution to a dynamically hotter distribution with more structure.

\begin{figure}
\plotone{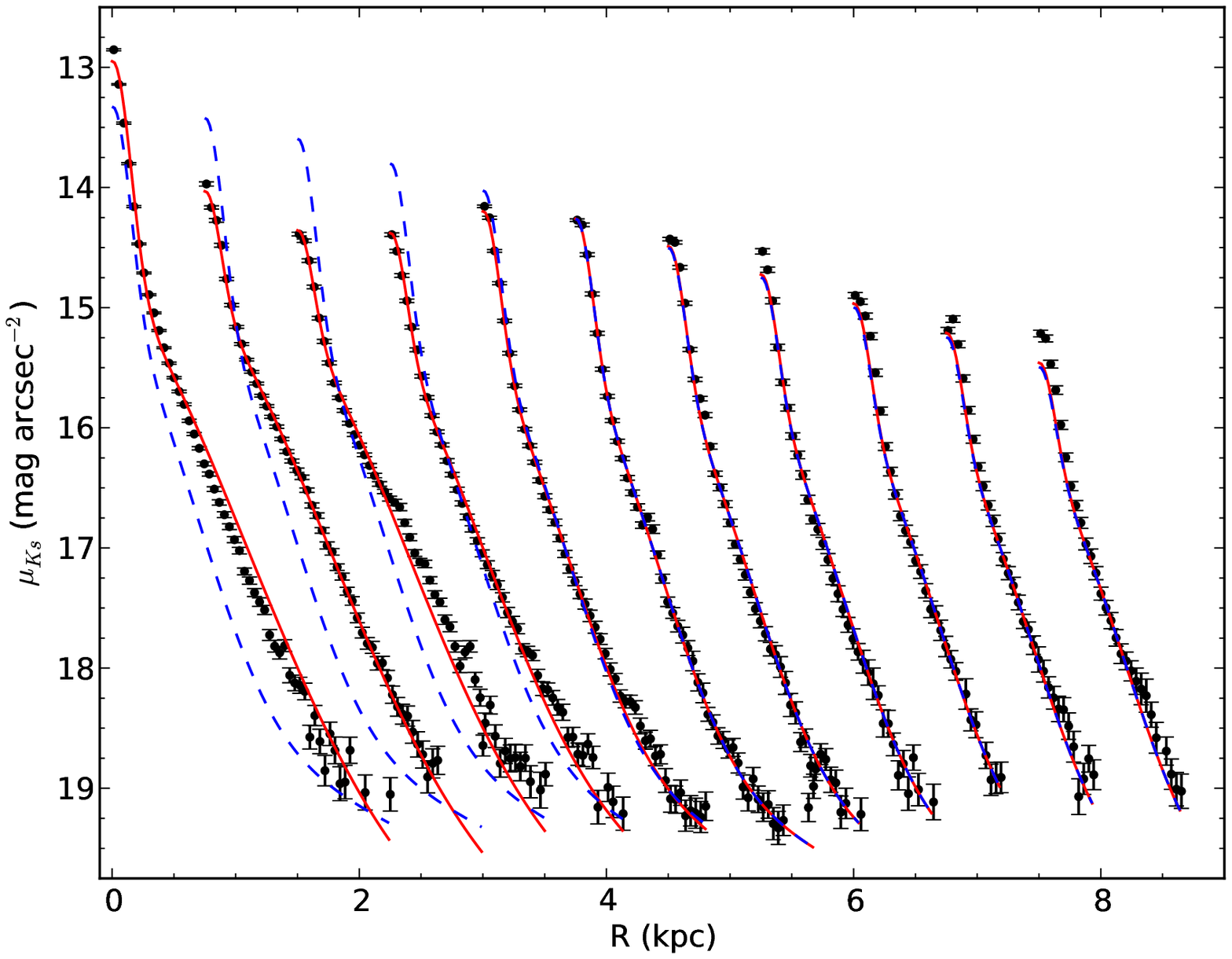}
\caption{Vertical surface brightness profiles of the attenuation corrected
  $\ks$-band data (black points). All four quadrants of the galaxy have been
  avaraged together. The leftmost point of each
  profile is at the radius of that profile, starting at R = 0 kpc and stepped
  every 0.75 kpc. Blue dashed lines show the best-fitting sech$^{2}$
  3-component disk model without any bulge/bar components, while solid red lines
  show the best-fitting sech$^{2}$ 3-component model with a truncation, a
 bar (approximated by a disk), and a nuclear super-thin disk.}
\label{fig:modelvertprofs}
\end{figure}

This transition is more
apparent as a function of radius in vertical profiles (Figure
\ref{fig:modelvertprofs}). While the 3-component model well fits the data
outside of 3 kpc, the data clearly has a dearth of flux inside of 3 kpc right
up until the very central regions of the galaxy. At small radii, however, there is a clear excess of flux at larger heights. We suggest this excess is due to a bar because previous studies have
found evidence for such a feature in NGC~891
(e.g. \citealt{Garcia-Burillo95}). Further, there are clearly kinks in the
vertical light profile (most evident at R=1.5 and z=0.8 kpc) caused by the `X'
shape of the projected bar \citep{Combes81,Mihos95,Bureau99,Bureau06}. Other
than these small kinks, however, the bar appears to be very close to having a pure
exponential vertical light distribution. 

Understanding the galaxy's behavior near the center is critical to estimating
the total luminosity of each disk component and to fully characterize their
radial distribution. Therefore we perform a second set of 2D fits on the data attenuation-corrected by Model B. While using the same basic scheme
as in Section \ref{sec:2dfitting}, this time we leave the central region of
NGC~891 unblanked. This necessitates several new model
parameters. We approximate the bar in edge-on projection as an outer-truncated
disk. This  appears to be
a good approximation to the light distribution of the bar seen edge-on, and
the functional form of a truncated disk simplifies our fitting procedure. Our
bar model therefore has three
free parameters: a central
SB ($\mu_{0,bar}$), scale-length ($h_{R,bar}$), and scale-height
($h_{z,bar}$). 

Since there is an excess of light at small R and z that appears
morphologically disconnected from the bar emission (see the top two panels of Figure \ref{fig:modelresid}), it
is necessary to introduce an additional component. We first attempted to fit a
classical $R^{1/4}$-law bulge to NGC~891, but found that this form made for
extremely poor fits. We found instead that the central
component of NGC~891 was well-fit by a nuclear disk with the same
scale-height as the super-thin disk. We therefore fixed the scale-height of
this component $h_{z,nuc}=h_{z,ST}$ and allowed the central SB $\mu_{0,nuc}$
and scale-length $h_{R,nuc}$ to vary. 

Finally, we add a truncation radius $R_{trunc}$, which functions as an
inner cutoff for all three main disks and an outer cutoff for the bar and
nuclear disk. For simplicity this truncation is abrupt: at all points inside
(outside) of
$R_{trunc}$ the main disks (bar and nuclear disk) have zero
flux. In reality there will be a transition length instead of a discontinuous
truncation, but we do not find that the data warrants this complication in the
model. The scale-heights of the three main disks should be well
determined by our original fitting scheme, and tests indicated that there were
minimal changes in central SBs and scale-lengths of the thin and thick disks
when the central region of NGC~891 was uncensored. Therefore we fixed the parameters of the thin and thick components at their best-fitting values in Table
\ref{tab:2dfit_newdust}, but we allowed $\mu_{0,ST}$ and $h_{R,ST}$ to
vary. We perform these new
fits only for sech$^{2}$ disks. 

\begin{deluxetable}{c c c c}
\tabletypesize{\footnotesize}
\tablewidth{0pt}
\tablecaption{Two-Dimensional sech$^{2}$ Fits, with Bar and Nuclear Disk}
\tablehead{\colhead{} & \multicolumn{2}{c}{Allowed Range} & \colhead{} \\
  \colhead{Parameter$^{\mathrm{a}}$} & \colhead{Min} & \colhead{Max} &
  \colhead{sech$^{2}$}}
\startdata
$\mu_{0,ST}$&25&10&13.39$\pm$0.02\\
$h_{R,ST}$&0.1&10&2.30$\pm$0.023\\
$h_{z,ST}^{\mathrm{b}}$&\nodata&\nodata&0.06\\
$L_{tot,ST}^{\mathrm{c}}$&\nodata&\nodata&4.04$\times 10^{10}$(25)\\
$\mu_{0,T}^{\mathrm{b}}$&\nodata&\nodata&15.32\\
$h_{R,T}^{\mathrm{b}}$&\nodata&\nodata&4.11\\
$h_{z,T}^{\mathrm{b}}$&\nodata&\nodata&0.25\\
$L_{tot,T}^{\mathrm{c}}$&\nodata&\nodata&6.86$\times 10^{10}$(42)\\
$\mu_{0,Th}^{\mathrm{b}}$&\nodata&\nodata&18.71\\
$h_{R,Th}^{\mathrm{b}}$&\nodata&\nodata&4.80\\
$h_{z,Th}^{\mathrm{b}}$&\nodata&\nodata&1.44\\
$L_{tot,Th}^{\mathrm{c}}$&\nodata&\nodata&2.13$\times 10^{10}$(13)\\
$\mu_{0,bar}$&25&10&15.42$\pm$0.01\\
$h_{R,bar}$&0.1&10&1.28$\pm$0.009\\
$h_{z,bar}$&0.01&2.5&0.38$\pm$0.001\\
$L_{tot,bar}^{\mathrm{c}}$&\nodata&\nodata&2.49$\times 10^{10}$(15)\\
$\mu_{0,nuc}$&25&10&13.23$\pm$0.04\\
$h_{R,nuc}$&0.1&10&0.25$\pm$0.007\\
$h_{z,nuc}^{\mathrm{b}}$&\nodata&\nodata&0.06\\
$L_{tot,nuc}^{\mathrm{c}}$&\nodata&\nodata&8.32$\times 10^{9}$(5)\\
$R_{trunc}$&1.5&4.5&3.09$\pm$0.014\\
$\Delta$z$^{\mathrm{b}}$&\nodata&\nodata&0.05\\
Reduced $\chi^{2}$&\nodata&\nodata&4.55\\
\enddata
\tablenotetext{a}{$\mu$ values in mag arcsec$^{-1}$, $h_{R}$, $h_z$,
  $\Delta$z, and $R_{trunc}$ values in kpc.}
\tablenotetext{b}{Values fixed using best-fitting values for the 3-component sech$^{2}$ model.}
\tablenotetext{c}{Total luminosities are in units of $L_{\odot,K}=3.909\times 10^{11} W Hz^{-1}$ (from
  \citealt{Binney98}). Values in parenthesis are the percentage of the total
  disk luminosity in that component.}
\label{tab:bulgefits}
\end{deluxetable}

The parameters of the best-fitting model are listed in Table
\ref{tab:bulgefits} and compared to the data and untruncated 3-component
sech$^{2}$ disk model (from Table \ref{tab:2dfit_newdust}) in Figure
\ref{fig:modelvertprofs}. The residual image from this
model is shown in the bottom panel of Figure \ref{fig:modelresid}. The outstanding residuals
that remain are associated with the `X' shape pattern attributed to the
bar. It would be very useful to have a simple analytic parameterization of
this light distribution, but we are not aware of one in the literature. While still not perfect, adding a
disk truncation, a
simple representation for the bar, and a nuclear disk dramatically improves our
ability to fit the inner portion of NGC~891. The central SB and scale-length
of the super-thin disk change minimally from the original fits. The bar
has a scale-height $\sim$1.5 times that of the thin disk, but substantially
smaller than the thick disk, which agrees with a visual inspection of our
images. The reduced $\chi^{2}$ of this fit is somewhat larger than for the
fits with only the three disk components; this is due to the unmodeled `X'
shape bar pattern.

\section{Discussion}
\label{sec:discussion}
\subsection{Literature Comparisons}
\subsubsection{Radial and vertical Disk Scale-lengths}
\label{sec:paramcomp}
Several groups have estimated NGC
891's thin disk parameters. Studies simultaneously fitting the thin and thick
disk components of NGC~891 typically find
optical thin disk scale-heights of $\sim$0.5 kpc (\citealt{Ibata09};
\citealt{Morrison97} and references therein). Using single component
exponential fits in the infrared, \citet{Aoki91}
determine a K-band scale-height of 0.35 kpc, while 
\citet{Xilouris99} find a K-band scale-height of 0.34 kpc and a K-band
scale-length of 3.87 kpc with their single-disk model. Due to the resolution
and depth of these studies these single-disk fits probe the region of the
surface-brightness profiles where the thin disk dominates, and thus are most
directly matched to this component when compared to multi-component fits. The decrease in scale-height in
the NIR may be due to the decreased role of dust attenuation. If vertical
profiles are fit to data that hasn't been corrected for attenuation the dust will
function to artificially increase the inferred scale-heights. Alternatively,
if vertical profile fits are restricted to sufficiently large z to avoid dust,
then the fit is biased toward the light distribution of the thick(er) disk
components. Supporting this hypothesis is the fact
that \citet{Xilouris99} find an optical scale-height of $\sim$0.4 kpc from
their RT models, closer to their NIR scale-heights than the $\sim$0.5 kpc
found in \citet{Ibata09} and \citet{Morrison97}.

 For our single component
exponential fits (the most directly comparable fits to most of the
above-mentioned literature) we obtain slightly smaller scale parameters (82\% smaller scale-height, 92\% smaller
scale-length) than \citet{Xilouris99}. Our increased resolution, combined with
the fact that the super-thin part of the vertical light distribution is at
high S/N, are the likely reasons for the decrease in scale-height of our
single component fits. The scale-length of the model with the
sech$^{2}$ vertical profile is very similar to the fit using the exponential distribution. This is unsurprising since regardless of the functional form
of the scale-height we use exponentials for the scale-length. The scale-height
of the sech$^{2}$ model, however, is only 68\% of the exponential
scale-height. Indeed, a comparison between the exponential and sech$^{2}$
scale-heights for all our models shows that the sech$^{2}$ scale-height is
always somewhat smaller than the exponential scale-height. As noted above, this discrepancy
arises naturally from the shallower nature of the sech$^{2}$ profile near the
midplane, which forces smaller scale-heights to be used to reproduce the
steeper 
features near the midplane.

For our multiple component fits, the thin disk is most closely approximated by the
second disk component, so we will refer to it as the `thin disk
component'. In our 2-component fits, the thin disk component has (except for
the sech$^{2}$ scale-height) larger scale parameter values than the Xilouris
K-band fits. This discrepancy seems likely to be due to the need for the thin
disk component to fit light from both the thin disk as well as the thick disk
(due to the depth of our observations),
while the light nearest to the midplane is fit well by the inner disk
component. Corroborating evidence for this hypothesis comes from the
3-component fits, which have thin disk component scale parameters very close to
the K-band values reported by \citet{Xilouris99}. 

In addition to the changing scale-heights for each disk of our 3-component
models, we also see a positive correlation in the scale-lengths of the
components. The scale-length of the super-thin disk for both the exponential
and sech$^{2}$ fit is $\sim$50\% of the scale-length of the thin disk, while
the thin disk scale-length is only $\sim$80\% of the size of the thick
disk. The scale-heights, however, still change more rapidly than the
scale-lengths. Defining $q\equiv h_{R}/h_{z}$, $q_{Th}\sim 3 < q_{T}\sim
15<q_{ST}\sim 30$. 

Of special interest in our 2- and 3-component models is the need for a disk
with a scale height even smaller than
that of the thin disk component of NGC~891. This component, which
we term the `super-thin disk component,' is usually hidden from view by the dust
lane as it has a scale-height of around 60-80 pc (perhaps slightly higher,
based on the results of the 2-component fits). Even prior
studies which attempted dust correction/modeling could not be expected to
reveal this disk in the NIR as it is very
close to even our own resolution limit ($\sim$40 pc), and certainly would be
unresolved in the data used by e.g. \citealt{Aoki91}. We note, however,
that \citet{Xilouris98} do require a super-thin disk to fit excess light in their
{\it optical} bandpasses (which have better resolution), with $h_{z}$=0.15 kpc
but an infinite scale-length with an arbitrary cutoff. With our data we are able to fit a non-infinite scale length, which we
find to be $\sim$2 kpc (again, slightly higher in the 2-component
fits). This is roughly half the size of the scale-lengths of the thin and
thick disk components, indicating that the stars in the super-thin disk are
very centrally concentrated compared to the stars in the other disks.

A likely conjecture is that
the super-thin disk in NGC~891 corresponds to the young star-forming disk seen
in the MW. The average scale-heights given by \citet{Bahcall80} for young
($\sim$90 pc) and old ($\sim$325 pc) stars in the MW correlate well with the best-fitting
parameters for NGC~891's super-thin and thin disk components. Before now this
disk has never been observed as a discrete vertical component in a galaxy
similar to the MW due to dust obscuration. However the presence of a
super-thin star-forming disk has been inferred from the inability of RT models to
reproduce the FIR SED of edge-on galaxies based only on known thin and thick
disk parameters
\citep{Popescu00,Bianchi08,MacLachlan11,Holwerda12}. In these models, the scale-height
of such a disk is generally assumed to either be similar to that of the MW
(e.g. \citealt{Popescu00}) or the same as the dust
(e.g. \citealt{Bianchi08}).  Our
unambiguous determination of the super-thin disk scale-height removes the need
to make such assumptions. Our finding that the super-thin disk scale-length is
significantly smaller than the thin or thick disk scale-length is also
important since previous studies have incorrectly assumed the super-thin disk
has the same scale-{\it length} as the thin disk.

\subsubsection{Disk Luminosities}
\label{sec:totlum}
For our single component fits (as well as for the thin disk component of our
3-component exponential fits) we compute comparable scale-lengths and
scale-heights but $\sim$1
mag arcsec$^{-2}$ brighter central surface brightnesses than found by \citet{Xilouris99}. This result implies that
our single component models predict $\sim$50\% higher total $\ks$-band
luminosities for NGC~891. Our images do go deeper than those of
\citet{Xilouris99}, which would boost the luminosity of our model. Our attenuation correction, which takes into account the
effects of clumpy dust, may also contribute to this discrepancy. We note that
our total luminosities are in much better agreement with more recent works
(e.g. \citealt{Popescu11}), which will be discussed in more detail later in this section.

For exponential and sech$^{2}$ disks the total flux can be computed by
integrating over R and z. Including the effect of an inner disk truncation
gives 
\begin{align}
F_{tot,exp} =
&2\pi h_{z}f_{\nu,0}10^{-\mu_{0}/2.5}e^{-R_{trunc}/h_{R}} \nonumber\\
&\times (h_{R}+R_{trunc})\\
F_{tot,sech^{2}} =
&4\pi h_{z}f_{\nu,0}10^{-\mu_{0}/2.5}e^{-R_{trunc}/h_{R}} \nonumber \\
&\times (h_{R}+R_{trunc})
\end{align}
for the super-thin, thin, and thick disks, where $R_{trunc}$ is the truncation
radius and $f_{\nu,0}$ is the flux zero-point. In this work we use the 2MASS value of 667
Jy for the $\ks$-band zero-point \citep{Cohen03}. The equations for the total flux of the bar and nuclear disk are
similar:
\begin{align}
F_{tot,exp} =
&2\pi h_{z}f_{\nu,0}10^{-\mu_{0}/2.5}(h_{R}-e^{-R_{trunc}/h_{R}} \nonumber\\
&\times (h_{R}+R_{trunc}))\\
F_{tot,sech^{2}} =
&4\pi h_{z}f_{\nu,0}10^{-\mu_{0}/2.5}(h_{R}-e^{-R_{trunc}/h_{R}} \nonumber\\
&\times (h_{R}+R_{trunc})).
\end{align}
We show the total luminosities of each component for both
3-component models as well as the model which adds a bar and nuclear disk
(the `truncated' model) 
in Tables \ref{tab:2dfit_newdust} and \ref{tab:bulgefits}.

For the truncated model we find that the bar contributes roughly the
same total luminosity as the thick disk, despite being confined to the
central regions of NGC~891. The nuclear disk is by far the faintest,
with a total luminosity only 40\% of the second least-luminous component (the
thick disk). The total $\ks$-band luminosity of the truncated model is
slightly ($\sim$5\%) smaller than the 3-component model. The fact that
the $\ks$-band light lost by truncating the disks of the 3-component
model is almost entirely made up for by the light from the bar and
nuclear disk is tantalizing evidence that disk instabilities have
merely rearranged the orbits of existing stars in NGC~891's inner
regions (e.g., \citealt{Combes81,Combes90,Raha91}). 

The super-thin disk has a total $\ks$-band
luminosity of $5.05\times 10^{10}$ ($6.62\times 10^{10}$) $L_{\odot,K}$ for the exponential (sech$^{2}$)
three disk model. The truncated super-thin disk has a total luminosity
$\sim$62\% of its non-truncated counterpart. Assuming the nuclear disk is an
inner continuation of the super-thin disk and adding its luminosity to that of
the truncated super-thin disk increases this ratio to $\sim$74\%. Assuming that the `young' stellar populations used as inputs to
RT models correspond exactly to our super-thin disk, the truncated disk luminosity is about
an order of magnitude larger than $5.36\times 10^{9}$ $L_{\odot,K}$, the value
assumed by \citet{Popescu11} for modeling the integrated SED of NGC~891, and
roughly 4 times larger than assumed for our RT modeling. We again note that due to
the nature of our attenuation correction Model B tends to have more luminous
central surface-brightnesses, but even the truncated total super-thin disk
luminosity of the Model A fits are still $\gtrsim$2 times as large
than the luminosity used in \citet{Popescu11}. These discrepancies are all
significantly larger than either our random or systematic errors, which are of
order $\sim$10\%.

However, when the {\it total} flux (bulge, young disk, and old disk) from the
model used by \citet{Popescu11} is compared to the total flux of our truncated
model, they agree to within 10\%. It therefore seems likely that the input
SEDs used by \citet{Popescu11} need to be adjusted, with the young stellar SED
getting boosted in the NIR at the expense of flux from the old stellar
SED. Adjusting the bolometric luminosities of the two components will not
solve this problem --- boosting the luminosity of the young component at all
wavelengths will cause an over-prediction of the UV flux (or an over-prediction
of the FIR light if the extra UV emission is suppressed by dust). It is plausible that
\citet{Popescu11}'s young stellar SED underestimates the impact of cool, evolved
stars, especially thermally-pulsating asymptotic giant branch stars which are
known to be difficult to model and are very bright in the NIR
\citep{Maraston05}. If this is the case it would explain why \citet{Popescu11}
are able to fit the UV and FIR flux of NGC~891 while under-predicting the NIR
flux of the young stellar disk component of the galaxy. Depending on the
spectral type of these cool stars the peak of the old component's SED will shift,
but it is not possible with our current data to predict exactly how these
input SEDs will change.

\citet{King90} report a B-band thin disk integrated luminosity of
6.7$\times 10^{9}$ $L_{\odot}$, which (even though it's in the optical) should be minimally affected by dust
because it is computed away from the midplane of NGC~891. Using our fitted thin-disk $\ks$-band
integrated luminosity, the thin disk has a B-$\ks$ color of $\sim$4.5 mag
(Vega), similar to a K4III \citep{Johnson66} star. Using the values from
what \citet{Popescu11} call the old disk, which is most similar to our
thin disk, gives a B-$\ks$ color of $\sim$5.2 mag, close to a
K5III or M0III. Based on a galactic evolution model \citet{Just96} find I-K
colors for the thin disks of edge-on galaxies to be $\sim$1.2 mag, similar to the colors of a K3III star. While these models were
not designed specifically for NGC~891, they are another
indication that the NIR luminosity in \citet{Popescu11} needs to be
redistributed between the young and old stars.

\subsection{Impact of the vertical fitting function}
\label{sec:expvssech2}

\subsubsection{The luminosity profile and total light}

As mentioned in Section \ref{sec:profiles}, the functional form of the
vertical light distribution has been much debated, with different
authors favoring different indices of $n$ in Equation
\ref{eq:vertprof}.  Due to dust obscuration near the midplane, shallow
image depth, or both, most of the prior studies have only been able to
accurately constrain fits with one (or at most two, when the data are
deep enough to probe the thick disk) disk component(s). However, our data
provides constraints for three disks, and while our 1- and 2-component
exponential models produce smaller reduced $\chi^{2}$ than the
equivalent sech$^{2}$ model, our 3-component model results in
sech$^{2}$ and exponential fits that are nearly indistinguishable from
each other. This similarity is also seen in the inferred total
luminosity ($L_{TOT}=L_{ST}+L_{T}+L_{Th}$) of the exponential model and
the sech$^{2}$ model, where $L_{ST}$, $L_{T}$, and $L_{Th}$ are the
super-thin, thin and thick components respectively. Despite differences
in individual component parameters
$\frac{L_{TOT,exp}}{L_{TOT,sech^{2}}} = 1.00$.  The same ratio for the
1- and 2-component models is $\sim$1.15 and $\sim$1.06, respectively.

These result indicates (i) that the presence of multiple disk
components must be considered when fitting the vertical light
distribution; and (ii) a distinction between sech$^{2}$ and
exponential fitting functions cannot be made in this context because
when adequate components are introduced to fit the observed light
distribution, the different functions are indistinguishable.

Due to the difficulty of statistically distinguishing between
different empirical forms of the vertical surface brightness
distribution, it could be argued that physically motivating a vertical
profile from equations of dynamical equilibrium would lead to more
accurate results. \citet{Comeron11b} take this approach in their
vertical surface brightness profile fitting, and find theoretical
values that seem to more closely resemble sech function, intermediate
in concentration at small z between exponential or sech$^{2}$
functions. To test the sech distribution we performed a 2D partially
fixed 3-component disk parameter fit to our $\ks$-band data (corrected
with RT Model B) using this distribution. We find that all fitted free
parameter values for the sech distribution fall between the fitted
values from the equivalent model using the exponential and sech$^{2}$
distributions (see the 12th and 14th columns of Table 
\ref{tab:2dfit_newdust}), with a reduced $\chi^{2}$ only 0.01
better than the corresponding sech$^{2}$ fit.

This result indicates that it is not necessary to compute dynamically
consistent profiles in order to parameterize the light distribution of
galaxy disks: empirical functions can be used with the same accuracy
provided there are adequate components.

\subsubsection{Component scales and scale ratios}

It still might be argued, however, that dynamically self-consistent
vertical profile functions would enable a more physical interpretation
of the derived parameters characterizing the light profile. On the
other hand, in light of \citet{Bovy12}'s argument that disk
components are likely continuous, extant dynamical analysis assuming
discrete components may be misleading. Nonetheless, the observed
vertical light profiles do appear to have clear breaks at 0.2 and
1.1 kpc, and this makes it reasonable to consider discrete components.
As such, it is important to understand the systematics attributable to
different choices for the vertical fitting function.

We find that for all of our models (1, 2, and 3-component) the cuspier
exponential vertical profiles yield larger scale-heights relative to
sech$^2$ vertical profiles. Although both models adopt a radially
exponential light distribution, models with exponential {\it vertical}
profiles yield smaller {\it radial} scale-lengths. Combined, the disk
oblateness changes by roughly a factor of 1.5 between models with
vertical light profiles that are exponential or sech$^2$, with disk
models becoming apparently thinner for sech$^2$ vertical
profiles. Again, this holds for all components in 1, 2, and
3-component models.

In the 3-component model the disk oblateness also changes more rapidly
between thin and thick components (a factor of $\sim$5) compared to
the oblateness change between super-thin and thin components (a factor
of $\sim$2).  These different oblateness changes are driven mostly by
changes in the radial scale-length, which increases by almost a factor
of 2 between super-thin and thin components, but only by $\sim$20\%
between thin and thick components. In contrast the scale-height ratio
for thin to super-thin components is $\frac{h_{z,T}}{h_{z,ST}} \sim
3.6 (4.2)$ for the exponential (sech$^{2}$) fits while for thick to
thin components the ratio is $\frac{h_{z,Th}}{h_{z,T}} \sim 5.0 (5.8)$
for the exponential (sech$^{2}$) fits.  Not surprisingly for our
2-component models the scale-height ratio for the two components is
intermediate between the results for thin-to-superthin and
thick-to-thin ratios.

The scale-height ratios for NGC~891 think-to-thin components are
comparable to those found by \citet{Comeron11b}, based on thin-thick disk
decompositions of edge-on spirals in the S$^{4}$G survey, for morphological type
T = 3 (Hyperleda indicates NGC~891 has T = 3.1$\pm$0.4). The good agreement of our results for
thin and thick disk thickness ratios to those of \citet{Comeron11b}, who
only fit two disks to their data, indicates either that the presence
of unaccounted-for flux from super-thin disks does not significantly
bias their results for the ratios of scale-heights of thin and thick
components or that most of their sample is without super-thin disk
components.

Changes in relative component luminosities are more subtle. For the ease of
the following disucssion we
tabulate the component luminosities of the exponential and sech$^{2}$ fits
with three disk components in Table \ref{tab:totlum}. For
3-component models, regardless of the functional forms of the vertical
light profile and whether there is a truncation, the thin component
contributes the most to the total disk luminosity, followed by the
super-thin component, and last the thick component. (In the
2-component model it is also the case that the second component, most
closely identified with the thin disk, also contributes the most
light.) If we take the difference between the exponential and
sech$^2$ values as characteristic of the systematic errors then we
find the three disk components, thin, super-thin and thick, contribute
51$\pm$4\%, 34$\pm$4\%, 15$\pm$1\%, respectively. The systematic error
in the thick-disk component luminosity is likely underestimated
because this component was fixed in scale-height and scale-length.

The most striking change in relative luminosities occurs between
superthin and thin disks in our three-component fits to NGC~891, with
the exponential model having $\frac{L_{ST}}{L_{T}}\sim 0.5$ compared to
$\frac{L_{ST}}{L_{T}}\sim 0.8$ for the sech$^{2}$ model. This comes about despite the
scale-height and scale-length ratios between superthin and thin
components being the same for both exponential and sech$^2$
models. What drives the difference in component luminosity is a change
in relative surface-brightness. In contrast to the superthin-to-thin
luminosity ratio, the luminosity ratios of thick-to-thin components is
$\frac{L_{Th}}{L_{T}}\sim 0.3$, i.e., essentially unchanged for
exponential, sech$^{2}$ or the truncated sech$^{2}$ models.

The relatively constancy of thick-to-thin luminosities is interesting
given that \citet{Comeron11b} report a systematic variation in
$\frac{L_{Th}}{L_{T}}$ when using different functional forms of the
vertical surface brightness profile, which they argue arises as a
result of how rounded (or concentrated) the model light profiles are
near the midplane. Specifically, they find thick-to-thin component
luminosity ratios 60\% larger than \citet{Yoachim06}. The latter adopt
a sech$^2$ vertical profile, less concentrated than that adopted by
\citet{Comeron11b}, which is closer to a sech function. The fact that
we obtain similar $\frac{L_{Th}}{L_{T}}$ ratios for both our fitting
functions might indicate that the addition of a super-thin disk
component removes this discrepancy. However, even for our 2-component
disk fits, the luminosity ratios only vary by 4\%, so something else
must be going on.

As an exercise, we took the sech$^2$ vertical profiles fit for
\citet{Yoachim06}'s sample, and fit to the sum of these components two
new 2-component models, one where both components have sech vertical
light profiles, and the other where both components have exponential
vertical light profiles. We found that the luminosity ratios for the
sech vertical profiles were roughly 60\% larger than those for the
input model sech$^2$ vertical profiles (the exponential vertical
profiles were 60\% larger again relative to the ratios for the sech
vertical profiles). This is the same factor found by
\citet{Comeron11b}. Both our exercise here and \citet{Comeron11b}'s
are one-dimensional fits (in z only). Our conclusion is that the
difference in thick-to-thin luminosity ratios is an artifact of
one-dimensional fitting which is largely eliminated when
two-dimensional light-profiles fits are performed.

\begin{deluxetable*}{ccccccc}
\tabletypesize{\footnotesize}
\tablewidth{0pt}
\tablecaption{Model B Three Disk Model Total Luminosities ($L/L_{\odot,K}$)$^{\mathrm{a}}$}
\tablehead{&\multicolumn{4}{c}{No Truncation}&\multicolumn{2}{c}{Truncated, with Bar + Nuclear Disk}\\
&\colhead{exp}&\colhead{\%$^{\mathrm{b}}$}&\colhead{sech$^{2}$}&\colhead{\%$^{\mathrm{b}}$}&\colhead{sech$^{2}$}&\colhead{\%$^{\mathrm{b}}$}}
\startdata
$L_{ST}^{\mathrm{c}}$&$5.05\times 10^{10}$&29&$6.62\times 10^{10}$&38&$4.04\times 10^{10}$&25 \\
$L_{T}^{\mathrm{c}}$&$9.60\times 10^{10}$&55&$8.41\times 10^{10}$&48&$6.86\times 10^{10}$&42 \\
$L_{Th}^{\mathrm{c}}$&$2.85\times 10^{10}$&16&$2.52\times 10^{10}$&14&$2.13\times 10^{10}$&13 \\
$L_{bar}$&\nodata&\nodata&\nodata&\nodata&$2.49\times 10^{10}$&15 \\
$L_{nuc}$&\nodata&\nodata&\nodata&\nodata&$8.32\times 10^{9}$&5 \\
$L_{tot}$&$1.75\times 10^{11}$&100&$1.76\times 10^{11}$&100&$1.64\times 10^{11}$&100 \\
\enddata
\tablenotetext{a}{$L_{\odot,K}=3.909\times 10^{11} W Hz^{-1}$ from
  \citet{Binney98}.}
\tablenotetext{b}{Percentage of total luminosity for each component.}
\tablenotetext{c}{ST, T, and Th are equivalent to disk components 1,2, and 3 in Table \ref{tab:2dfit_newdust}.}
\label{tab:totlum}
\end{deluxetable*}

\subsection{MIR Vertical Profiles}
Of value to our investigation would be corroborating evidence of a super-thin
disk in NGC~891 in another dataset, ideally in the MIR which has been claimed
to be ideal for studies of starlight with minimal attenuation \citep{Sheth10}. Towards that
end we analyzed the archival IRAC 4.5$\mu$m data looking for a super-thin
component consistent with our NIR fits. While the IRAC data has resolution
only half that of the lowest resolution of our WHIRC data, we note that our
$\ks$-band fits indicate that NGC~891's super-thin disk has a central
surface brightness $\sim$1.5-2 mag arcsec$^{-2}$ brighter than its thin disk,
and $\sim$4 mag arcsec$^{-2}$ brighter than its thick disk. At 4 kpc in radius
and at the approximate
resolution of the IRAC data in height (1.8"$\sim$80 pc), the super-thin
disk surface brightness should be comparable to the fitted thin disk
surface brightness ($\sim$0.2 mag fainter for the exponential fit, and
$\sim$0.9 mag brighter for the sech$^{2}$ fit, assuming no color gradients
between $\ks$ and 4.5$\mu$m). Therefore it is not
unreasonable to expect to see at least some evidence of the super-thin disk in
the IRAC data, even though it may never be the dominant component in the
ranges we can probe.

\begin{figure*}
  \plotone{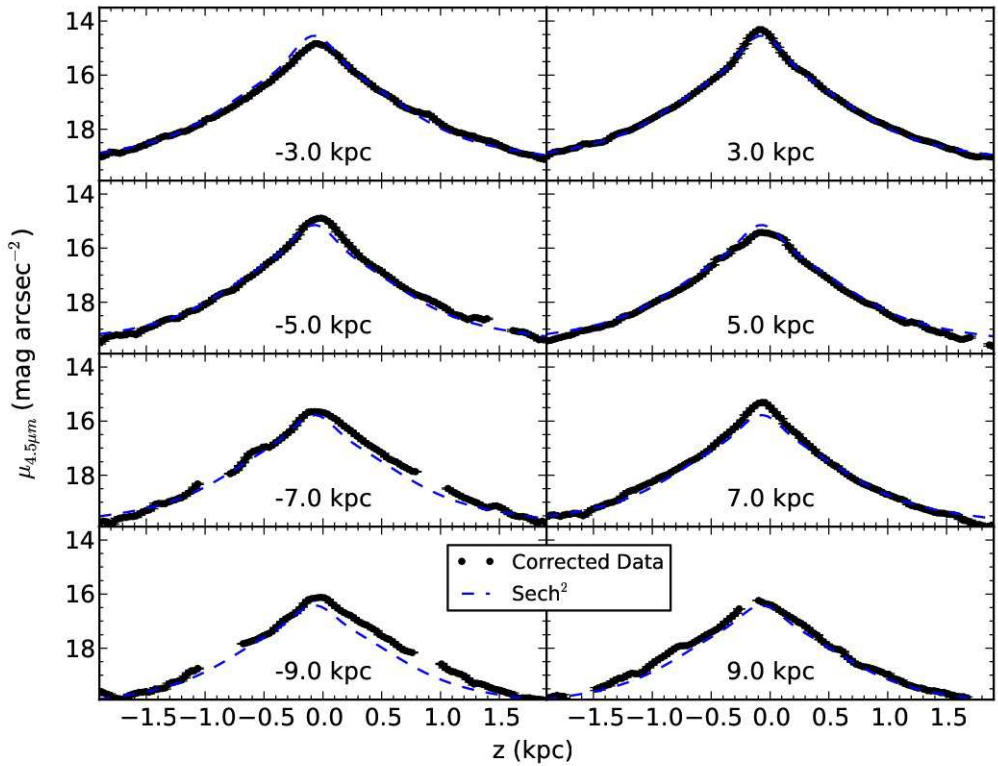}
  \caption{Three-component sech$^{2}$ model vertical
    profiles (blue lines), overlaid on the IRAC 4.5$\mu$m
    data (black points), at several radii. Gaps in the data are due to foreground stars which
    were removed before plotting. There are more gaps than for the WHIRC data
    (Figures \ref{fig:1compslice}-\ref{fig:3compslice}) due to a combination
    of very red stars and the increased size of the IRAC point-source
    function, which spreads more light from nearby stars near (but not in) our
  apertures into the profiles. Positive radii are on the SW side of the galaxy.}
\label{fig:1dirac}
\end{figure*}

We computed 2D fits to the IRAC 4.5$\mu$m data, using the same methodology as our
fits of the $\ks$-band WHIRC data, with a 3-component disk model. The exponential disk model had difficulty converging on
a physically plausible fit, due to degeneracies introduced by the lowered
resolution. The sech$^{2}$ model, however, produced a good fit (Table
\ref{tab:2dfit_IRAC} and Figure \ref{fig:1dirac}) with parameters very similar
to that for the $\ks$-band data. The scale-heights of
the super-thin and thin disk in the fits to the 4.5$\mu$m data are identical to the $\ks$-band values. 

In hindsight it is perhaps unsurprising that we were able to identify NGC~891's
super-thin disk at 4.5$\mu$m: \citet{Comeron11b} set out to decompose the
galaxies in the S$^{4}$G sample into thin and thick disk components, but several
(e.g. IC 2058, NGC~1495, and NGC~5470) have fits that are closer to what would
be expected for a super-thin+thin disk. This suggests not only that the IRAC
bands are capable of identifying super-thin disks in (very) nearby galaxies, but
that such disks may be common in the local universe. Attempting a 3-component decomposition of these S$^{4}$G galaxies would be a logical next step in
determining if they are truly super-thin+thin+thick disk
systems. Indeed, \citet{Comeron11a} perform a 3-component (1D) analysis on S$^{4}$G data of
NGC~4013, and find a system that appears to have a MW-like super-thin disk but
extended (in height) thin and thick disks. \citet{Comeron11a} do
not identify the narrowest disk as a super-thin disk but rather term the
components thin, thick, and `extended'; this difference is semantic until
there is a clear astrophysical distinction between the components (e.g. age).

Although the lower-resolution IRAC data appears good enough to identify
super-thin disks in at least some nearby edge-on galaxies there are still
several advantages to our methodology. First and
foremost is the increased resolution in the NIR, which allows super-thin
disks to be constrained at much larger distances. This is especially important
for the study of MW-sized galaxies, which are relatively rare in the local
universe. The increased resolution is helpful even in nearby galaxies: lower
spatial resolution increases the likelihood of
degeneracies in the fits, which is already known to be an important issue in
performing edge-on disk decompositions \citep{Morrison97}. 
The primary motivation for undertaking vertical disk decompositions in the MIR
is the limited dust attenuation at those wavelengths and the availability of
deep imaging data from {\it Spitzer}. However, our radiative
transfer models indicate there is as much as 0.5 mag of attenuation at
4.5$\mu$m (Figures \ref{fig:radialSBdustcompare} and \ref{fig:radialSBdustcompare_iraccorr}). While most of the 4.5$\mu$m light experiences
very little attenuation given the concentration of dust near the midplane, a
full, accurate accounting of the MIR light will {\bf still} require
an attenuation correction. 

Further, by using $\ks$-band data we ensure that we exclusively fit the light from the stellar
disk(s) and avoid contamination from dust, which \citet{Meidt12} find to
be $\ge$5\% of total integrated light and up to 20\% of the light locally from
star-forming regions in spiral galaxies. While MIR dust emission does affect our attenuation
correction in the $\ks$-band by skewing the $\ks$-4.5$\mu$m color, this is a
second-order effect --- a correction to a correction in the $\ks$-band itself. The ability of our model to reproduce the data at a variety of
radii and heights (Figure \ref{fig:cc45profiles}) indicates that it has been
properly accounted for in our modeling. In contrast, dust emission is both a
first and second order correction at 3.6 and 4.5$\mu$m.

\begin{deluxetable}{cc}
\tabletypesize{\footnotesize}
\tablewidth{0pt}
\tablecaption{IRAC 4.5$\mu$m Model Parameters}
\tablehead{ \colhead{Parameter$^{\mathrm{a}}$}& \colhead{sech$^{2}$}}
\startdata
$\mu_{0,1}$&13.97$\pm$0.032\\
$h_{R,1}$&2.39$\pm$0.016\\
$h_{z,1}$&0.06$\pm$0.002\\
$\mu_{0,2}$&15.21$\pm$0.012\\
$h_{R,2}$&3.50$\pm$0.007\\
$h_{z,2}$&0.25$\pm$0.001\\
$\mu_{0,3}$&18.26$\pm$0.003\\
$h_{R,3}$&4.80$^{\mathrm{b}}$\\
$h_{z,3}$&1.44$^{\mathrm{b}}$\\
\enddata
\tablenotetext{a}{$\mu$ values in mag arcsec$^{-1}$, $h_{R}$, $h_z$ values in kpc.}
\tablenotetext{b}{Fixed values from \citet{Ibata09}.}
\label{tab:2dfit_IRAC}
\end{deluxetable}

Finally, there is additional value in making the attenuation correction in the
NIR as it opens up this important spectral region for future imaging and
spectroscopic efforts. For instance, there are several molecular band-heads
between 1 and 2.3$\mu$m which may be able to separate different types of cold
giant stars \citep{Lancon07}, allowing for age and composition gradients to be
determined in the stellar disks. 

\section{Conclusion}
\label{sec:conclusion}
We have obtained sub-arcsecond resolution NIR imaging of NGC~891, a nearby
massive spiral galaxy frequently compared to the Milky Way, using the WHIRC
instrument on the WIYN telescope. These images cover roughly $\pm$10.5 kpc in
radius and all relevant regions for vertical disk decompositions. We
reduced the data using our own pipeline, specifically designed for the
reduction of images with very extended sources. We supplemented this data with
archival {\it Spitzer} IRAC 3.6 and 4.5$\mu$m data, primarily to assist in computing an
attenuation correction.

Concurrently, we produced a 3D Monte Carlo RT SED model of NGC~891
using parameters known to fit the UV-FIR SED of the galaxy. A comparison of
the IR colors predicted by the model with our WHIRC and IRAC data indicated that the balance of dust grain sizes was
different near the midplane compared to at large heights. We therefore
constructed a second model with
a lowered PAH fraction near the midplane, which provided better fits to the
data at small heights while leaving the color distribution at large heights unchanged. Both of these models
were used to construct an attenuation correction based on the $\ks$-4.5
$\mu$m color. The difference in the attenuation correction between the two
models is small (to first order the difference starts at 0 mag, increases at a
rate of $\sim$0.4 mag per mag of $\ks$-band attenuation up to $\aeks$=1.5 mag,
then increases at a much smaller rate of $\sim$ 0.07 mag per magnitude of attenuation) and does not significantly affect the following results. 

Using a Levenberg-Marquardt nonlinear curve fitting routine, we fit 2D
seeing-convolved surface-brightness profiles to the $\ks$-band data, using models
with up to three disk components. The vertical light distributions of the disks were parameterized by either
exponential or sech$^{2}$ functions, while the radial distributions were
assumed to be exponential. It is immediately clear that single disk component
models provide poor fits to the data. Two- and 3-component models do much
better, and a close inspection of the vertical profiles shows that
three components are necessary to fully model the light distribution. The
3-component model produces a {\it thin} disk with scale-length ($h_{R}$)
$\sim$4 kpc and scale-height ($h_{z}$) $\sim$0.27 kpc, and a {\it super-thin}
disk with $h_{R}$ and $h_{z}$ of $\sim$2.1 kpc and $\sim$0.07 kpc  respectively. The
thin and super-thin disks are consistent with prior work on NGC~891 and the
MW's young, star-forming disk (albeit with a smaller scale-length). We find
that as $h_{z}$ decreases from the thick to the super-thin disk the oblateness
$\left(\frac{h_{R}}{h_{z}}\right)$ increases from $\sim$3 to $\sim$30, even though
$h_{R}$ is also decreasing. Without data sensitive to
star-formation rate we cannot
definitively state that NGC~891's super-thin disk is star-forming like its putative
Milky Way counterpart, but this does
seem to be the most likely possibility, particularly given the evidence for
embedded star-formation from the observed FIR excess in this galaxy. 

While our 3-component models fit the data well outside of the central bulge
region, near the center of NGC~891 this model produces significant residuals
with respect to the data,
overestimating the light near the midplane while underestimating light away
from the midplane. These residuals are consistent with what would be expected
from a bar. CO
observations have indicated that NGC~891 has a bar, and the presence of such a
structure is clear from X-shaped isophotes visible in our data,
which are characteristic of bar orbits. The simplest explanation consistent with the
data is that NGC~891's bar truncates the galaxy's main disk components and
redistributes the stars into the bar structure. Other than the relatively low
surface-brightness X-shaped isophotes the bar appears to be well fit by a disk
with a scale-length of $\sim$1.3 kpc and a scale-height of $\sim$0.4 kpc.

We also find excess light at the very center of NGC~891. This excess
is well fit by a nuclear disk with the same scale-height as the super-thin
disk, a scale-length of 0.25 kpc ($\sim$10\% of the super-thin disk), and 20\% of the super-thin disk's luminosity; a classical R$^{1/4}$-law bulge cannot
adequately fit this feature and in fact this type of spheroid does not appear to be present at all in NGC
891. 

A truncation between the central components (the nuclear disk and bar) and the
three main disks is also necessary to accurately model the entire $\ks$-band
light distribution of NGC~891. This truncation happens at roughly 3.1 kpc. New models with this nuclear disk, a rough approximation to the bar, and a disk/bar truncation are able to produce good
fits to our data all the way down to the center of NGC~891.

We find that when three
disk components are modeled, both exponential and sech$^{2}$ vertical
distribution functions fit the data equally well, even at our sub-arcsecond  resolution. Despite the apparent difference in individual
fitted parameters between the sech$^{2}$ and exponential 3-component disk
model (e.g. sech$^{2}$ disks tend to have smaller scale-heights than
exponential disks), we find that they predict nearly identical {\it total} $\ks$-band
luminosities for NGC~891 (models with fewer components predict slightly larger
total luminosities for the exponential disks, on the order of $\sim$10\%). The sech$^{2}$ disk model including the bar, nuclear disk, and
disk/bar truncation has comparable total luminosity,
$\sim$95\% as bright as the simpler 3-component disk model with no truncation. Based on the
truncated model we find that the super-thin component (including the nuclear
disk) contributes
roughly 30\% of the total $\ks$-band luminosity in NGC~891, while the thin disk,
thick disk, and bar respectively contain $\sim$42\%, 13\%, and 15\% of the total. The luminosity and scale height ratios of NGC~891's thin and thick disk agree
with those found for other edge-on spirals in the S$^{4}$G study of comparable
type and rotation speed, even though the S$^{4}$G analysis only fits two disks
to their data.

The $\ks$-band luminosity of NGC~891's super-thin disk is a factor of 4-10 larger
than what has been adopted in radiative transfer models capable of reproducing
the UV-FIR integrated SED of
the galaxy \citep{Bianchi08,Popescu11}. The {\it total} disk luminosity is, however, very similar between
these models and our results. The most likely explanation for this discrepancy
is that some of the NIR light that had been attributed to the thin+thick disks in
these earlier models actually belongs to the super-thin component. Prior to
the present work NGC~891's super-thin disk had never been directly detected,
and although its UV luminosity could be inferred by using energy-balance
arguments, there were no direct constraints on its luminosity in the NIR or
its spatial location. Because the UV luminosity of the super-thin disk {\it
  is} fairly well constrained, the additional NIR light seen in our data
cannot be accounted for by uniformly increasing the intrinsic super-thin disk
luminosity at all wavelengths. Rather, the NIR luminosity must be boosted
without affecting the intrinsic super-thin disk SED in the UV, likely through
an increased amount of cool evolved stars, such as RSGs and AGBs. Confirmation
of this conjecture will be the topic of a future paper.

Finally, we determine that there is also an excess of light near NGC~891's
midplane in IRAC 4.5$\mu$m data, with scale height comparable to what we find
for the super-thin disk in the NIR. However, while useful for verifying the
validity of our methodology, there are several reasons why IRAC data are
non-ideal for identifying super-thin stellar disks. First, contemporary NIR
instruments yield much higher angular resolution than their MIR counterparts; NIR detectors will be able to probe super-thin disks at larger distances
from the MW. Secondly, we have shown that the IRAC bands have up to $\sim$0.5
mag of dust attenuation near the midplane on NGC~891. Therefore, any attempt
to determine the intrinsic light profile from a spiral galaxy's disk at low
heights will need to employ an attenuation correction like the one derived in
this work, even in the MIR. Finally, both the IRAC 3.6 and 4.5$\mu$m filters
(the ones most commonly used to measure stellar flux) have non-negligible
amounts of non-stellar emission from PAH and hot dust particles. Any attempt
to compute the stellar flux at these wavelengths must therefore correct for
this emission, which is known to be highly variable across individual
galaxies. By combining our NIR observations with existing MIR data, we are able to probe
deeper than if we used each wavelength regime independently: We retain the
spatial resolution and minimal susceptibility to contamination from dust
emission of the NIR, while producing excellent attenuation corrections using
the combination of the NIR and the MIR colors.

\acknowledgements{This research was supported by NSF AST-1009471. We thank Arthur Eigenbrot for assistance with the
  observations, Dick Joyce for useful conversations about
  optimizing WHIRC efficiency, and Jeff Percival for help with creating the
  dither scripts used for the observations. We also acknowledge Tom
  Robitaille, Barb Whitney,
  and Kenny Wood for assistance with the radiative transfer modeling and the
  dust physics. Discussions with Bob Benjamin were very useful in comparing
  the disk components to those in the MW. We also thank Blakesley Burkhart and
  Corey Wood for helpful edits and revisions. Comments from an anonymous
  referee were also very useful in improving this work. We
  acknowledge the usage of the HyperLeda database
  (http://leda.univ-lyon1.fr). This work is based in part on observations made with the Spitzer Space Telescope, obtained from the NASA/ IPAC Infrared Science Archive, both of which are operated by the Jet Propulsion Laboratory, California Institute of Technology under a contract with the National Aeronautics and Space Administration.}

\appendix
\section{WHIRC Data Reduction Procedure}
\label{sec:rdx}
\subsection{Initial Data Inspection}
Before (and during) the data reduction process we inspected the individual
images for artifacts and defects. Three images from both the SW and NE $\ks$-band pointings
contained elevated signal levels over a large region near the left-middle of
the detector and were removed from the reduction. Additionally, during a few
exposures of the central $\ks$-band and SW H-band the sky background appears to
undergo a rapid change, faster than our sky-source duty cycle. We therefore
also removed four SW H-band exposures and five central $\ks$-band exposures from
the reduction pipeline.

\subsection{Individual Image Reduction}
\label{sec:datared}
Data reduction generally followed the procedures described in the WHIRC Quick
Guide to Data Reduction
manual\footnote{http://www.noao.edu/kpno/manuals/whirc/WHIRC\_Datared\_090824.pdf},
although we modified several steps to optimize the reduction for very extended
targets like NGC~891. We detail all reduction steps here for clarity, going
into depth for the non-standard portions of our reduction.

First, excess data on the images were trimmed off. Then, images taken with
Fowler-4 sampling (J-band) were divided by four to remove the additive nature
of the sampling. We then applied a linearity correction to all images with the
IRAF task IRLINCOR using coefficients provided in the data reduction
manual. The coordinate matrix of the image headers was then adjusted to take
into account the rotation of the telescope during the observations.

We took 20 dome flat exposures for each filter: 10 with the dome lamps on and
10 with the lamps off. The 10 lamp-on flat images were scaled to
have the same mean flux in the central quarter of the images and then averaged
together to produce a combined lamp-on flat; this process was then repeated
with the lamp-off flats. The combined lamp-off flat was
then subtracted from the combined lamp-on flat to remove any signal not coming
from the flat lamps (e.g. dark current, scattered light, or thermal emission in the dome).

Before reducing the flat-fields any further, a first-order
correction must be applied because WHIRC has a `pupil ghost', likely due to internal reflections
in the imager's optics. It appears in the central 80\arcsec of the detector.
this feature is most prominent in the $\ks$-band, with an amplitude about 15\%
of the background, but it is also present (albeit at much lower
levels) in
J and H as well. The ghost was modeled and removed from the flat-fields (it is
removed from the object frames naturally during the sky-subtraction process) using the representative pupil
template provided on the WHIRC
web page\footnote{http://www.noao.edu/kpno/manuals/whirc/datared.html} and the
IRAF MSCRED.RMPUPIL task. While pupil templates can be constructed from J and
$\ks$ flats, for our purposes the provided template was more than
adequate, with errors $<$0.3\% on the corrected images. We note that the flats we used are from an earlier night in our run
(UT date 2011 Oct 16); we found night-to-night deviations in the flat-fields of
$<$1\% over the entire image area. All object and sky images were flattened
before proceeding with further data processing. 

Finally, the central quarter of the pupil-ghost-removed flat was normalized to an average value of 1. Small and negative pixel
values (caused by bad pixels and later masked out) were replaced by unity, to prevent erroneously large
and/or negative flux values from appearing in the flat-fielded science
images. 
\subsection{Sky Subtraction}
With extended sources like NGC~891, sky subtraction can be very difficult. To
combat this problem, we interleaved observations of a relatively empty sky
field $\sim$10\arcmin\,away (with similar dither amplitudes between sky images as for the data) with our observations of NGC~891. For each set of on-source dithers (4
for J and H, 9 for $\ks$) we combine sky images taken before (2 for J and H, 5
for $\ks$) and after (again, 2 for J and H and 5 for $\ks$) to produce a sky frame
suitable for use on the data images. Each sky frame was scaled to have the
same median value to avoid biasing the statistics. The sky frames were then median combined iteratively pixel-by-pixel with 2$\sigma$ clipping until the median converged
or all the pixels in the individual sky frames had been clipped out. 

After
all pixels have been median combined, we used an iterative relaxation algorithm
(described in Appendix \ref{sec:relaxation}) to interpolate sky
values for sky image pixels where all the constituent pixels were clipped
out. The relaxation procedure provides accurate interpolation and converges
very quickly when the pixels needing interpolation are relatively few and
widely spaced; we deliberately chose to obtain dedicated sky images and
ensured that dither amplitudes were larger than the stellar
point-source-functions (PSFs) in order to make this process as efficient and accurate
as possible. For a given set of on-source images the appropriate combined
sky frame was scaled to have the same median value as the data frame and then
subtracted. The resulting sky-subtracted image was then field flattened by the
normalized flat-field image.

Prior to registration, the sky-subtracted data images were cleaned of (most) chip
defects using the IRAF task FIXPIX (a linear interpolator) and a pixel mask created from the ratio of two lamp-on
flats with different exposure times. The data images were then distortion
corrected using the IRAF task GEOTRAN and distortion files downloaded from the
WHIRC web page. 

WHIRC has a very small dark current ($\sim$0.2 DN
sec$^{-1}$ pix$^{-1}$) compared to the sky background in most filters. For
comparison, WHIRC J-band has an average sky brightness of $\sim$6 DN
sec$^{-1}$ pix$^{-1}$. (Dark current is not an issue even in principle
  for dome flats since we utilize `on-off' differences.) Additionally, the sky subtraction procedure functions
to remove most of the (already small) contribution to the image flux from the
dark current. However, the dark current exhibits a fixed pattern structure. Further, the correct procedure for dark current removal is subtraction
{\it before} field-flattening. Nonetheless, due to our limited set of dark
frames, we were unable to demonstrate an improvement in image flatness with
dark subtraction. Therefore a separate dark subtraction has minimal impact on the
reduced images. For future studies focusing on low surface-brightness
extended structure we do recommend a more thorough investigation of
dark-current subtraction.

When compared to a super-sky frame made up of an average of a given night's
sky images, the dome flats appear to have small (on the order of 1\%)
illumination errors. However, the illumination appears to vary between nights
and potentially even during nights, making correction difficult. We tested an
illumination correction on a small subset of our data and found it made very
little difference in the reduced images. Due to the limited improvement and
variability of the illumination we do not employ an illumination
correction, but again we recommend a more careful exploration of this issue
for detailed low surface-brightness studies.

\subsection{Relative Image Registration and Mosaics}
Because of WHIRC's relatively small field of view and the extremely extended
nature of NGC~891 on the chip, fully automatic registration procedures
(e.g. SCAMP; \citealt{Bertin06}) were found to be unreliable. Therefore we
performed relative image registration between individual exposures by hand,
first computing source catalogs for each image using Source Extractor
\citep{Bertin96}, then hand-identifying individual sources that overlapped
between images. Based on visual comparisons between individual registered
images and the PSF of the final mosaics, our registration scheme is accurate
to $\lesssim$1 pixel.

With the images all registered in relation to each other, we created full J, H, and
$\ks$ mosaics using the Swarp software package \citep{Bertin02}. Because of the extended nature of NGC
891 compared to the WHIRC field-of-view the  estimate of the average flux in
the individual
data frames used to subtract the combined sky frame is biased, and therefore
the background of the sky-subtracted images is over-subtracted. To combat this,
we used the high S/N image mosaics to hand-produce very conservative masks of NGC~891 (and
foreground stars), essentially masking out all pixels except those very close
to the edge of the mosaic where galaxy contamination is minimized. This global
mask was then converted into a pixel mask for each individual frame. We then
re-ran the pipeline, starting from the sky-subtraction step, this time
normalizing the sky images to the source images using only unmasked
pixels. The relative WCS offsets were held constant from the first iteration
of the pipeline. This two-step process effectively removes the
over-subtraction bias from the extended nature of NGC~891 in our images.

\subsection{Absolute WCS Registration and Flux Calibration}
The next step was to register the image mosaics in an absolute astrometric system. First we hand-constructed catalogs of
the centroids of stars visible on both the WHIRC $\ks$-band image and the IRAC
4.5 $\mu$m image (see Section \ref{sec:iracdata} for details on the IRAC
images) using Ds9 \citep{Joye03}. Using the IRAC image as an astrometric
reference is useful not only because we will be using $\ks$-IRAC colors in our
analysis and therefore want good registration between these datasets but also
because both the IRAC and WHIRC data go much deeper than standard NIR
calibration catalogs. We then used the IRAF task CCMAP to perform the
registration. To preserve the excellent relative agreement between bands the
$\ks$-band absolute
calibration was applied manually to the J and H bands. 

To flux calibrate the images, stars from the 2MASS PSC with reliable photometry
were matched to Source Extractor integrated fluxes on the mosaics. We then fit
the relationship between log(DN) and Vega magnitude, obtaining fits
with standard errors of 0.02 (0.03, 0.02) mag in the J (H, $\ks$)
bands. 

\section{Iterative Relaxation Algorithm}
\label{sec:relaxation}
When stacking dithered images together to produce a sky frame, the ideal
situation is to have enough exposures so that every pixel on the stacked
frame has
a large number of samples of the sky background with which to do statistics. In reality, however, 
even for relatively sparse fields it is not uncommon to find a few pixels on
the stacked frame which were masked out on all of the input individual
images, usually due to foreground or background contaminants appearing at
multiple dither positions. For these pixels the sky background must be
interpolated. As long as this happens only to a
small fraction of the total field, we can still construct a very accurate sky
background using our iterative relaxation method, described below. 

For each `problematic' pixel we first find the nearest pixel with a valid sky
value, and use that value as an initial guess. Then, the average value of all
adjacent pixels (including bad pixels) is computed, with pixels diagonally adjacent given weights of
1/$\sqrt{2}$ the weights of horizontally and vertically adjacent pixels. The
weighted average then becomes the next guess for the pixel value, and the
difference between the initial guess and the weighted average is
computed. This process iterates for all problematic pixels until the average
change per pixel between iterations is below 1\%. When the bad pixels are
surrounded (or nearly surrounded) by good pixels, this process converges in
just a few iterations. While our dither strategy minimizes source overlap on
multiple dithers, it is not perfect and there are a few regions with multiple
adjacent bad pixels. Generally, however, convergence occurs
in under 100 iterations. 

\section{Attenuation Fitting Scheme and Formulae}
\label{sec:attenuationdetails}
To construct an attenuation correction we fit a curve to the projected image pixels
of the RT model. Because the attenuation curve is relatively complex, we
employ a piecewise approach to accurately fit the attenuation without relying
on very high-order functions, which can have undesirable edge behavior.

For $\ks$-4.5$\mu$m $\lesssim$ 1.5 mag we find that a fourth order polynomial
fits the model trends of $\aeff$ versus color quite well. For larger values of
the attenuation we approximate the relationship as a line. To avoid any bias near the endpoints of the polynomial
fit, we join the line to the polynomial where $\ks$-4.5$\mu$m = 1.3 mag. To
connect the two functions we require that both the value and the first
derivative of both functions be equal at $\ks$-4.5$\mu$m = 1.3 mag. This ensures
that the attenuation correction is continuous at all points. 

For Model A, we fit 
\begin{eqnarray}
\aej &=& \left\{
  \begin{array}{lr}
    -0.03 + 2.58(\ks-4.5\mu m) + 0.19(\ks-4.5\mu m)^2&\\- 0.67(\ks-4.5\mu m)^3 +
    0.23(\ks-4.5\mu m)^4 & : (\ks-4.5\mu m) \le 1.3\\
    1.70(\ks-4.5\mu m) + 0.62 & : (\ks-4.5\mu m) > 1.3
  \end{array}
\right.\\
\aeh &=& \left\{
  \begin{array}{lr}
    -0.01 + 1.62(\ks-4.5\mu m) + 0.27(\ks-4.5\mu m)^2&\\- 0.30(\ks-4.5\mu m)^3 +
    0.10(\ks-4.5\mu m)^4 & : (\ks-4.5\mu m) \le 1.3\\
    1.67(\ks-4.5\mu m) - 0.01 & : (\ks-4.5\mu m) > 1.3
  \end{array}
\right.\\
\aeks &=& \left\{
  \begin{array}{lr}
    -0.01 + 1.04(\ks-4.5\mu m) + 0.18(\ks-4.5\mu m)^2&\\- 0.10(\ks-4.5\mu m)^3 +
    0.05(\ks-4.5\mu m)^4 & : (\ks-4.5\mu m) \le 1.3\\
    1.42(\ks-4.5\mu m) - 0.28 & : (\ks-4.5\mu m) > 1.3.
  \end{array}
\right.
\end{eqnarray}
For Model B, we obtain
\begin{eqnarray}
\aej &=& \left\{
  \begin{array}{lr}
    0.70(\ks-4.5\mu m) + 5.90(\ks-4.5\mu m)^2&\\- 5.59(\ks-4.5\mu m)^3 +
    1.59(\ks-4.5\mu m)^4 & : (\ks-4.5\mu m) \le 1.3\\
    1.73(\ks-4.5\mu m) + 0.92 & : (\ks-4.5\mu m) > 1.3
  \end{array}
\right.\\
\aeh &=& \left\{
  \begin{array}{lr}
    -0.01 + 0.34(\ks-4.5\mu m) + 3.93(\ks-4.5\mu m)^2&\\- 3.15(\ks-4.5\mu m)^3 +
    0.82(\ks-4.5\mu m)^4 & : (\ks-4.5\mu m) \le 1.3\\
    1.75(\ks-4.5\mu m) - 0.22 & : (\ks-4.5\mu m) > 1.3
  \end{array}
\right.\\
\aeks &=& \left\{
  \begin{array}{lr}
    0.19(\ks-4.5\mu m) + 2.50(\ks-4.5\mu m)^2&\\- 1.72(\ks-4.5\mu m)^3 +
    0.40(\ks-4.5\mu m)^4 & : (\ks-4.5\mu m) \le 1.3\\
    1.50(\ks-4.5\mu m) - 0.10 & : (\ks-4.5\mu m) > 1.3.
  \end{array}
\right.
\end{eqnarray}

\end{document}